\documentclass[twocolumn,showpacs,preprintnumbers,amsmath,amssymb,superscriptaddress, prl,floatfix]{revtex4-1}
\usepackage{bm}
\usepackage{graphicx}
\usepackage{dcolumn}
\usepackage{tabularx}
\usepackage{bm}
\usepackage{epstopdf}
\usepackage{amsmath}
\usepackage{color}
\usepackage{here}
\begin{document}

\preprint{Preprint}

\title{Spin anisotropic interactions of lower polaritons in the vicinity of polaritonic Feshbach resonance}

\author{N. Takemura}
\email[E-mail: ]{naotomot@phys.ethz.ch}
\affiliation{Laboratory of Quantum Optoelectronics, Institute of Physics, \'Ecole Polytechnique F\'ed\'erale de Lausanne, CH-1015, Lausanne, Switzerland}
\author{M. D. Anderson}
\affiliation{Laboratory of Quantum Optoelectronics, Institute of Physics, \'Ecole Polytechnique F\'ed\'erale de Lausanne, CH-1015, Lausanne, Switzerland}
\author{M. Navadeh-Toupchi}
\affiliation{Laboratory of Quantum Optoelectronics, Institute of Physics, \'Ecole Polytechnique F\'ed\'erale de Lausanne, CH-1015, Lausanne, Switzerland}
\author{D. Y. Oberli}
\affiliation{Laboratory of Quantum Optoelectronics, Institute of Physics, \'Ecole Polytechnique F\'ed\'erale de Lausanne, CH-1015, Lausanne, Switzerland}
\author{M. T. Portella-Oberli}
\affiliation{Laboratory of Quantum Optoelectronics, Institute of Physics, \'Ecole Polytechnique F\'ed\'erale de Lausanne, CH-1015, Lausanne, Switzerland}
\author{B. Deveaud}
\affiliation{Laboratory of Quantum Optoelectronics, Institute of Physics, \'Ecole Polytechnique F\'ed\'erale de Lausanne, CH-1015, Lausanne, Switzerland}

\date{\today}

\begin{abstract}
We determine experimentally the spinor interaction constants of lower polaritons $\alpha_1$ and $\alpha_2$ using a resonant pump-probe spectroscopy with a spectrally narrow pump pulse. Our experimental findings are analyzed with the Bogoliubov-type theory and a mean-field two channel model based on the lower polariton and biexciton basis. We find an enhancement of the attractive interaction and a dissipative non-linearity of lower polaritons with anti-parallel spins in the vicinity of the biexciton resonance when the energy of two lower polaritons approaches energetically that of the biexciton. These observations are consistent with the existence of a scattering resonance between lower polaritons and biexcitons (polaritonic Feshbach resonance).     
\end{abstract}


\maketitle
\section{Introduction}
In semiconductor microcavities, the strong coupling between quantum well excitons and cavity photons gives rise to new eigenstates: lower and upper polariton. \cite{Weisbuch1992}. In the last two decades, exciton-polaritons have drawn a large attention of physicists in various fields both in terms of fundamental physics and potential device applications. Among them, recent achievements are  realization of Bose-Einstein condensation \cite{Kasprzak2006} and polariton superfluidity \cite{Amo2009,Utsunomiya2008,Kohnle2011,Kohnle2012}. In these physics, the non-linearity originating from exciton-exciton interaction plays a central role. Moreover, a spin anisotropy of the exciton interactions emerges from the optically coupled two spin projections of a heavy-hole exciton: $|J_z=\pm 1\rangle=|J_e=\mp 1/2,J_h=\pm 3/2\rangle$. In particular, the interaction between excitons with anti-parallel spins allows the formation of a molecular
bound state of two excitons, called a biexciton \cite{Borri2000}. In cold atom physics, it is well known that the molecular bound state dramatically modifies inter atomic interactions when the energy of incident atoms in an open channel is tuned to that of a molecular state in a closed channel. This is a manifestation of a scattering resonance called Feshbach resonance \cite{Inouye1998}. Actually, the biexciton also affects the interaction of polaritons when the lower polariton energy is in the vicinity of the biexciton energy \cite{Wouters2007,Carusotto2010}. Experimentally, the signature is an enhancement of the interaction and a large polariton decay associated with a biexciton channel \cite{Vladimirova2010,Takemura2014,PhysRevB.91.201304}. These observations can be understood as a polaritonic analog of the Feshbach resonance: ``polaritonic Feshbach resonance" \cite{Takemura2014}. For device applications of semiconductor microcavities, the enhancement of the non-linearity might be useful for a non-classical photon generation \cite{Carusotto2010,Oka2008}. Additionally, the polaritonic Feshbach resonance is also interesting in terms of some questions of fundamental physics including a collective paring \cite{Marchetti2014} and a formation of bose polarons \cite{Casteels2014}. Moreover, the spin anisotropic polariton interaction is important in regard to the stability of polariton condensates. For instance, a thermodynamical argument predicts a collapse of polariton condensates when the polariton interaction with anti-parallel spins is stronger than that with parallel spins \cite{0268-1242-25-1-013001,Vladimirova2010}.

This paper explores the spinor interactions of the two lower polaritons in the vicinity of the polaritonic Feshbach resonance. Experimentally, we employ a polarization selective pump-probe spectroscopy with a spectrally narrow pump pulse. Depending on a polarization configuration of two pulses (cocirculary and countercirculary), we can independently access the polariton interaction with parallel ($\alpha_1$) and anti-parallel spins ($\alpha_2$). The central idea of the spectroscopic technique is similar to that of our previous paper\cite{Takemura2014,Takemura2014a}, in which the lower- and upper polariton branches were excited simultaneously with a spectrally broad pump pulse. In this paper, however, the pump pulse is a spectrally narrow and excites only the lower polariton branch. This allows the use of the lower polariton basis in a theoretical model, which dramatically simplifies the analysis by removing complicated effects associated with the upper polariton branch such as an excitation induced dephasing (EID) \cite{Takemura2016}. We discuss a quasi-stationary behavior of the spinor polariton physics within the Bogoliubov theory and a mean-field two-channel model.  

This paper is organized as following. Section II describes the experimental configuration. Section III presents the theoretical background used in our paper. In Section IV, we discuss the experimental results and analyze them. We present conclusions in the last section.

\section{Experiment}
\begin{figure}
\includegraphics[width=0.5\textwidth]{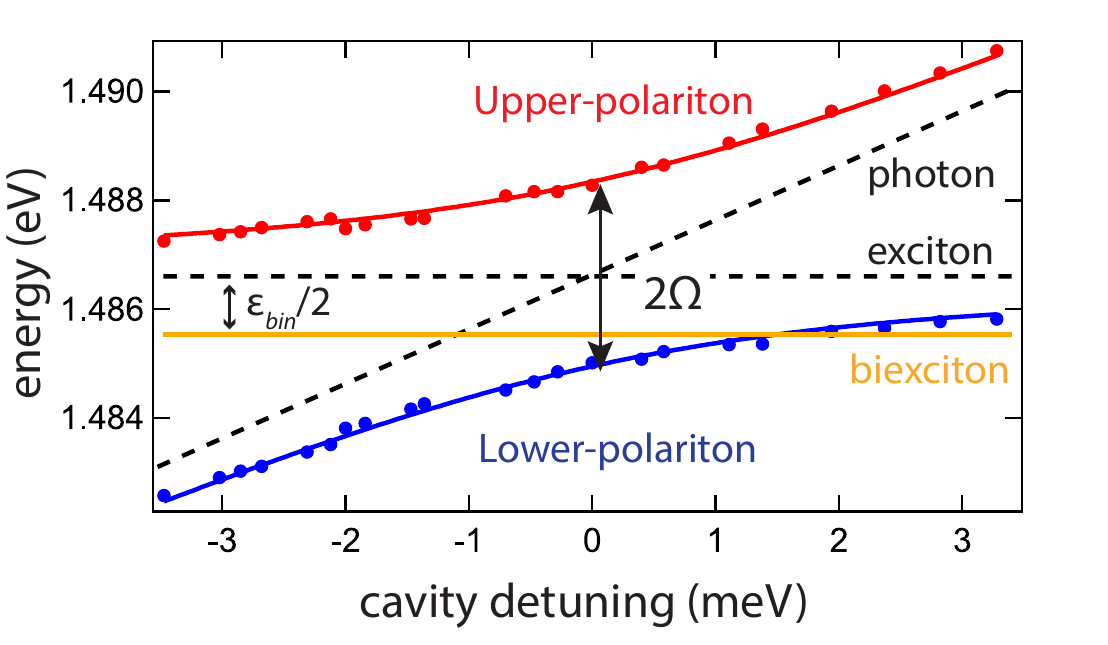}
\caption{Energies of the lower and the upper polaritons as a function of the cavity detuning $\delta$ at a zero in-plane momentum. The energy splitting between the two branches at zero cavity detuning is $2\Omega=$3.45 meV. The orange line represents the energy of the biexciton represented as $\varepsilon_x-\varepsilon_{bin}/2$, where $\varepsilon_{bin}$ is the binding energy of a biexciton}
\label{fig:detuning}
\end{figure}

The measurements are performed with a high quality III-V GaAs-based microcavity \cite{Stanley1994} at the cryogenic temperature of 4K. A single 8 nm In$_{0.04}$Ga$_{0.96}$As quantum well is embedded between a pair of GaAs/AlAs distributed Bragg-reflectors (DBR). The Rabi splitting energy is $2\Omega$=3.45 meV at a zero cavity detuning \cite{Kohnle2011}. Due to the wedge of the cavity spacer, we can tune the cavity detuning by moving the laser spot over the surface of the sample. Here, the cavity detuning is defined as $\delta=\varepsilon_c-\varepsilon_x$, where $\varepsilon_c$ and $\varepsilon_x$ are respectively the energies of the cavity photon and exciton modes at $\bm k= 0$. Depending on the cavity detuning, the energy of the lower polariton varies following the equation:
\begin{equation}
\varepsilon_{L}=\frac{1}{2}(\varepsilon_{c}+\varepsilon_{x})-\frac{1}{2}\sqrt{(\epsilon_{c}-\epsilon_{x})^2+(2\Omega)^2}
\label{eq:LPenergy}
\end{equation}
The change of the energy of the lower and upper polaritons is shown in Fig. \ref{fig:detuning} as a function of the cavity detuning. Figure. \ref{fig:detuning} shows that by sweeping the cavity detuning, the energy of the two lower polaritons can cross the biexciton energy.

The sample is excited with a strong pump pulse and probed with a weak probe pulse. Both pump and probe pulses are obtained from a broadband few hundreds femtosecond Ti:Sapphire laser (the linewidth is around 15 meV). After their separation, only the pump pulse is spectrally narrowed with a pulse shaper. The line width of the pump pulse is less than 1 meV. The pump pulse is tuned to the lower polariton energy; thus, it only excites the lower polariton branch. The in-plane momentum of the pump pulse is set at $\bm k= 0$ $\mu$m$^{-1}$. On the other hand, the probe pulse probes both lower- and upper polariton branches. The center of the probe spectrum is set between the two branches with an in-plane momentum ${\bm k}\sim 0$ $\mu$m$^{-1}$. Polarizations of the two pulses are either $\sigma^+$ or $\sigma^-$. Cocirculary and countercircular polarization pump-probe configurations produce polaritons with parallel and anti-parallel spins, respectively. The measurements are performed in a transmission configuration. For the detection of the probe pulse, we employ a heterodyne detection technique, which largely suppresses incoherent stray lights and greatly increases the signal to noise ratio \cite{Takemura2014a}. Additionally, we removed the noise originating from the laser spectrum envelope with a numerical low-pass filter. For all experiments, the pump and probe photon densities are respectively fixed as 15$\times$10$^{12}$ photons/pulse/cm$^{2}$ (1 mW) and 3.7$\times$10$^{12}$ photons/pulse/cm$^{2}$ (250 $\mu$W). Since the probe pulse is spectrally much broader than the pump, the density produced by the probe pulse is indeed much weaker than that by the pump pulse within a single polariton branch. In the main text, all pump-probe measurements are performed at zero pump-probe delay. The pump-probe time delay dependent spectra are discussed in Appendix A. 

\section{Spinor Gross-Pitaevskii equations}
In this section, starting from an exciton-photon basis Hamiltonian, we introduce a lower polariton basis Hamiltonian\cite{Rochat2000,Ciuti2000,Kuwata-Gonokami2000,Wouters2007} and derive lower polariton spinor Gross-Pitaevskii equations. The lower polariton spinor Gross-Pitaevskii equations is then used for analyzing pump-probe spectra.

\subsection{Exciton-photon Hamiltonian}
The Hamiltonian written in exciton, photon, and biexciton basis reads:
\begin{equation}
\hat{H}=\hat{H}_{\rm lin}+\hat{H}_{\rm int}+\hat{H}_{\rm qm}.
\label{eq:ecHamiltonian}
\end{equation}
This Hamiltonian is composed of a linear part: 
\begin{eqnarray}
\hat{H}_{\rm lin}&=&\sum_{\sigma=\left\{\uparrow\downarrow\right\}}\int d{\bf x} \Bigl[\hat{\bm \psi}_{x,\sigma}^{\dagger}(\varepsilon_{x}-\frac{\hbar^2\nabla^2}{2m_{x}})\hat{\bm \psi}_{x,\sigma}\nonumber\\
& &+\hat{\bm \psi}_{c,\sigma}^{\dagger}(\varepsilon_{c}-\frac{\hbar^2\nabla^2}{2m_{c}})\hat{\bm \psi}_{c,\sigma}\nonumber\\
& &+\Omega(\hat{\bm \psi}_{c,\sigma}^{\dagger}\hat{\bm \psi}_{x,\sigma}+\hat{\bm \psi}_{x,\sigma}^{\dagger}\hat{\bm \psi}_{c,\sigma})\Bigr]\nonumber\\
& &+\int d{\bf x}\ \hat{\bm \psi}_{B}^{\dagger}(\varepsilon_{B}-\frac{\hbar^2\nabla^2}{2m_{B}})\hat{\bm \psi}_{B},
\end{eqnarray}
an interacting part:
\begin{eqnarray}
\hat{H}_{\rm int}&=&\sum_{\sigma=\left\{\uparrow\downarrow\right\}}\int d{\bf x}\Bigl[\frac{1}{2}g\hat{\bm \psi}_{x,\sigma}^{\dagger}\hat{\bm \psi}_{x,\sigma}^{\dagger}\hat{\bm \psi}_{x,\sigma}\hat{\bm \psi}_{x,\sigma}\nonumber\\
& &+\frac{1}{2}g^{\scriptscriptstyle +-}\hat{\bm \psi}_{x,\sigma}^{\dagger}\hat{\bm \psi}_{x,-\sigma}^{\dagger}\hat{\bm \psi}_{x,-\sigma}\hat{\bm \psi}_{x,\sigma}\nonumber\\
& &+\frac{1}{2}g_{bx}(\hat{\bm \psi}_{B}\hat{\bm \psi}_{x,\sigma}^{\dagger}\hat{\bm \psi}_{x,-\sigma}^{\dagger}+\hat{\bm \psi}_{x,\sigma}\hat{\bm \psi}_{x,-\sigma}\hat{\bm \psi}_{B}^{\dagger})\nonumber\\
& &-g_{\rm pae}(\hat{\bm \psi}_{c,\sigma}^{\dagger}\hat{\bm \psi}_{x,\sigma}^{\dagger}\hat{\bm \psi}_{x,\sigma}\hat{\bm \psi}_{x,\sigma}+\hat{\bm \psi}_{x,\sigma}^{\dagger}\hat{\bm \psi}_{x,\sigma}^{\dagger}\hat{\bm \psi}_{x,\sigma}\hat{\bm \psi}_{c,\sigma})\Bigr],\nonumber\\
\end{eqnarray}
and a quasi-mode coupling to the electric field outside the cavity:
\begin{equation}
\hat{H}_{\rm qm}=\sum_{\sigma=\left\{\uparrow\downarrow\right\}}\int d{\bf x}\ \Omega_{qm}(\hat{\bm \psi}_{c,\sigma}^{\dagger}{\bm F}_\sigma+{\bm F}_\sigma^{*}\hat{\bm \psi}_{c,\sigma}).
\end{equation}
Here, $\Omega$ is the Rabi coupling between excitons and photons. ${\bm F}_\sigma$ represents the electric field outside the cavity, while $\Omega_{qm}$ is the strength of the quasi-mode coupling. The operator $\hat{\bm \psi}_{x(c),\sigma}$ is a bosonic exciton (photon) operator including  spins (circular polarization $\sigma^{\pm}$). The masses of excitons, photons, and biexcitons are respectively expressed as $m_x$, $m_c$, and $m_B$. The energies $\varepsilon_x$, $\varepsilon_c$, and $\varepsilon_B$ are respectively the energies of exciton, photon and biexciton with zero momentum. The spin-dependent exciton operators satisfy the following commutation relations:
\begin{equation}
[\hat{\psi}_{x(c),\sigma}({\bf x}),\hat{\psi}_{x(c),\sigma'}^{\dagger}({\bf x'})]=\delta_{\sigma,\sigma'}\delta({\bf x}-{\bf x'})
\end{equation}
and
\begin{equation}
[\hat{\psi}_{x(c),\sigma}^{(\dagger)}({\bf x}),\hat{\psi}_{x(c),\sigma'}^{(\dagger)}({\bf x'})]=0.
\end{equation} 
The bosonic biexciton operator $\hat{\bm \psi}_B$ satisfies:
\begin{equation}
[\hat{\psi}_{B}({\bf x}),\hat{\psi}_{B}^{\dagger}({\bf x'})]=\delta({\bf x}-{\bf x'})
\end{equation}
and
\begin{equation}
[\hat{\psi}_{B}^{(\dagger)}({\bf x}),\hat{\psi}_{B}^{(\dagger)}({\bf x'})]=0.
\end{equation}
In this Hamiltonian, the constants $g$ and $g^{+-}$ respectively represent the effective contact-type exciton-exciton interaction with parallel and anti-parallel spins. We would like to stress the difference between bare non-contact exciton-exciton interaction and an effective contact interaction employed here \cite{Proukakis1998}. The term $g_{+-}$ does not exist within the Born approximation, but appears only when higher-order scattering contributions are included. Actually, we will find that this ``background interaction" $g_{+-}$ is important to explain our experimental observations. In Appendix B, we discuss how the effective interactions are related to the bare non-contact exciton-exciton interactions. The term $g_{bx}$ is the exciton-biexciton coupling strengh. Finally, the term $g_{\rm pae}$ is a photon assisted exchange scattering \cite{Combescot2007,Ciuti2000,PhysRevB.80.155306}, which originates from the fermionic constituents of an exciton. This interaction is also often referred to as an anharmonic saturation term or ``phase-space filling" \cite{Ciuti2000}. In comparing with the experimental results, we will determine the values of the interaction constants. The extracted values will be presented in Table. \ref{table:parameters} in Section IV.    

\subsection{lower polariton Hamiltonian}
Now, we introduce the polariton basis $\hat{\bm \psi}_{L(U),\sigma}$ defined as  
\begin{eqnarray}
\left( \begin{array}{c}
\hat{\psi}_{x,\sigma}({\bf x}) \\
\hat{\psi}_{c,\sigma}({\bf x}) \\
\end{array} \right)
=
\left( \begin{array}{cc}
X_{\bm 0} & -C_{\bm 0} \\
C_{\bm 0} & \ X_{\bm 0} 
\end{array} \right)
\left( \begin{array}{c}
\hat{\psi}_{L,\sigma}({\bf x}) \\
\hat{\psi}_{U,\sigma}({\bf x}) \\
\end{array} \right),
\end{eqnarray}
the Hamiltonian is rewritten with the polariton basis. The coefficients $X_{\bm 0}$ and $C_{\bm 0}$ are the Hopfield coefficients at ${\bm k}=0$. They are respectively written as\cite{Deng2010} 
\begin{equation}
X_{{\bm 0}}=\sqrt{\frac{1}{2}\left(1+\frac{(\varepsilon_{c}-\varepsilon_{x})}{\sqrt{(\varepsilon_{c}-\varepsilon_{x})^2+(2\Omega)^2}}\right)}
\end{equation}
and
\begin{equation}
C_{{\bm 0}}=-\sqrt{\frac{1}{2}\left(1-\frac{(\varepsilon_{c}-\varepsilon_{x})}{\sqrt{(\varepsilon_{c}-\varepsilon_{x})^2+(2\Omega)^2}}\right)}.
\end{equation}
The field operators of the polariton also satisfy the bosonic commutation relations:
\begin{equation}
[\hat{\psi}_{L(U),\sigma}({\bf x}),\hat{\psi}_{L(U),\sigma'}^{\dagger}({\bf x'})]=\delta_{\sigma,\sigma'}\delta({\bf x}-{\bf x'})
\end{equation}
and
\begin{equation}
[\hat{\psi}_{L(U),\sigma}^{(\dagger)}({\bf x}),\hat{\psi}_{L(U),\sigma'}^{(\dagger)}({\bf x'})]=0.
\end{equation} 
Since the pump excites only the lower polariton, we neglect the terms involving the upper polariton operators. Using only the lower polariton and biexciton operators, the Hamiltonian Eq. \ref{eq:ecHamiltonian} is approximated as  
\begin{equation}
\hat{H}\simeq\hat{H}_{\rm lin, LP}+\hat{H}_{\rm int, LP}+\hat{H}_{\rm qm, LP},
\label{eq:LPHamiltonian}
\end{equation}
where $\hat{H}_{\rm lin, LP}$, $\hat{H}_{\rm int, LP}$, and $\hat{H}_{\rm qm, LP}$ are respectively 
\begin{eqnarray}
\hat{H}_{\rm lin,LP}&=&\sum_{\sigma=\left\{\uparrow\downarrow\right\}}\int d{\bf x}\ \hat{\bm \psi}_{L,\sigma}^{\dagger}(\varepsilon_L-\frac{\hbar^2{\nabla}^2}{2m_L})\hat{\bm \psi}_{L,\sigma}\nonumber\\
& &+\int d{\bf x}\ \hat{\bm \psi}_{B}^{\dagger}(\varepsilon_B-\frac{\hbar^2{\nabla}^2}{2m_B})\hat{\bm \psi}_{B},
\label{eq:HlintLP}
\end{eqnarray}
\begin{eqnarray}
\hat{H}_{\rm int,LP}&=&\sum_{\sigma=\left\{\uparrow\downarrow\right\}}\int d{\bf x}\left[\right.\frac{1}{2}\alpha_1\hat{\bm \psi}_{L,\sigma}^{\dagger}\hat{\bm \psi}_{L,\sigma}^{\dagger}\hat{\bm \psi}_{L,\sigma}\hat{\bm \psi}_{L,\sigma}\nonumber\\
& &+\frac{1}{2}g^{\scriptscriptstyle +-}_L\hat{\bm \psi}_{L,\sigma}^{\dagger}\hat{\bm \psi}_{L,-\sigma}^{\dagger}\hat{\bm \psi}_{L,-\sigma}\hat{\bm \psi}_{L,\sigma}\nonumber\\
& &+\frac{1}{2}g^{bx}_{L}(\hat{\bm \psi}_{B}\hat{\bm \psi}_{L,\sigma}^{\dagger}\hat{\bm \psi}_{L,-\sigma}^{\dagger}+\hat{\bm \psi}_{L,\sigma}\hat{\bm \psi}_{L,-\sigma}\hat{\bm \psi}_{B}^{\dagger})\left.\right],\nonumber\\
\label{eq:HintLP}
\end{eqnarray}
and
\begin{eqnarray}
\hat{H}_{\rm qm,LP}&=&\sum_{\sigma=\left\{\uparrow\downarrow\right\}}\int d{\bf x}\ \Omega_{L,0}(\hat{\bm \psi}_{L,\sigma}^{\dagger}{\bm F}_{\sigma}+{\bm F}_{\sigma}^{*}\hat{\bm \psi}_{L,\sigma}).\nonumber\\
\end{eqnarray}
Here, the energy $\varepsilon_L$ is given by Eq. \ref{eq:LPenergy}. The effective polariton mass
$m_{L}$ is obtained as $1/m_L=|X_{\bm 0}|^2/m_x+|C_{\bm 0}|^2/m_c$\cite{Deng2010} . The interaction constants $\alpha_1$, $g^{\scriptscriptstyle +-}_L$, and $g^{bx}_{L}$ are respectively defined as
\begin{equation}
\alpha_1=gX_{\bm 0}^4+4g_{\rm pae}X_{\bm 0}^3|C_{\bm 0}|,
\label{eq:alpha1}
\end{equation}
\begin{equation}
g^{\scriptscriptstyle +-}_L=g^{\scriptscriptstyle +-}X_{\bm 0}^4,
\end{equation}
and
\begin{equation}
g^{bx}_{L}=g_{bx}X_{\bm 0}^2.
\end{equation}
The quasi-mode coupling in the polariton basis is expressed as 
\begin{equation}
\Omega_{L,{\bm 0}}=\Omega_{qm}C_{\bm 0}.
\end{equation}
This Hamiltonian is similar to the one of the two-channel model in atomic Bose-Einstein condensates, which describes the Feshbach resonance \cite{Timmermans1999,Timmermans1999a}. 

\subsection{Spinor Gross-Pitaevskii of lower polariton and biexciton}
We derive the equations of motion of the wave functions of lower polaritons and of biexcitons from the Hamiltonian written in Eq. \ref{eq:LPHamiltonian}. Combined with a mean-field approximation and an inclusion of phenomenological decay rates of lower polaritons $\gamma_L$ and biexcitons $\gamma_B$, the Heisenberg equations of motion leads to two coupled equations:   
\begin{eqnarray}
i\hbar\dot{\psi}_{L,\sigma}({\bf x},t)&=&\Bigl[\varepsilon_L-i\gamma_L-\frac{\hbar^2\nabla^2}{2m_L}+\alpha_1|\psi_{L,\sigma}({\bf x},t)|^2\nonumber\\
& &+g^{\scriptscriptstyle +-}_L|\psi_{L,-\sigma}({\bf x},t)|^2\Bigr]\psi_{L,\sigma}({\bf x},t)\nonumber\\
& &+g^{bx}_{L}\psi_{B}({\bf x},t)\psi_{L,-\sigma}^*({\bf x},t)+f_{ext,\sigma}({\bf x},t)\nonumber\\
\label{GP_LPcr}
\end{eqnarray}
and
\begin{eqnarray}
i\hbar\dot{\psi}_{B}({\bf x},t)&=&\Bigl[\varepsilon_B-i\gamma_B-\frac{\hbar^2\nabla^2}{2m_B}\Bigr]\psi_{B}({\bf x},t)\nonumber\\
& &+g^{bx}_{L}\psi_{L,\uparrow}({\bf x},t)\psi_{L,\downarrow}({\bf x},t),
\label{GP_BX}
\end{eqnarray}
where the driving term $f_{ext,\sigma}$ is defined as  $f_{ext,\sigma}({\bf x},t)=\Omega_{L,{\bm 0}}F_\sigma({\bf x},t)$. The mean fields of the lower polariton and biexciton states are respectively represented as ${\psi}_{L,\sigma}=\langle\hat{\psi}_{L,\sigma}\rangle$ and ${\psi}_{B}=\langle\hat{\psi}_{B}\rangle$.
The above set of equations is similar to those that describe real time evolutions of the Feshbach molecule in atomic Bose-Einstein condensates \cite{Mackie2002,Timmermans1999,Timmermans1999a}. We note that similar equations are employed for the analysis of non-linear optical signals in exciton-biexciton coupled systems in quantum wells \cite{Fu1997,Neukirch2000}.  
\begin{figure}
\includegraphics[width=0.5\textwidth]{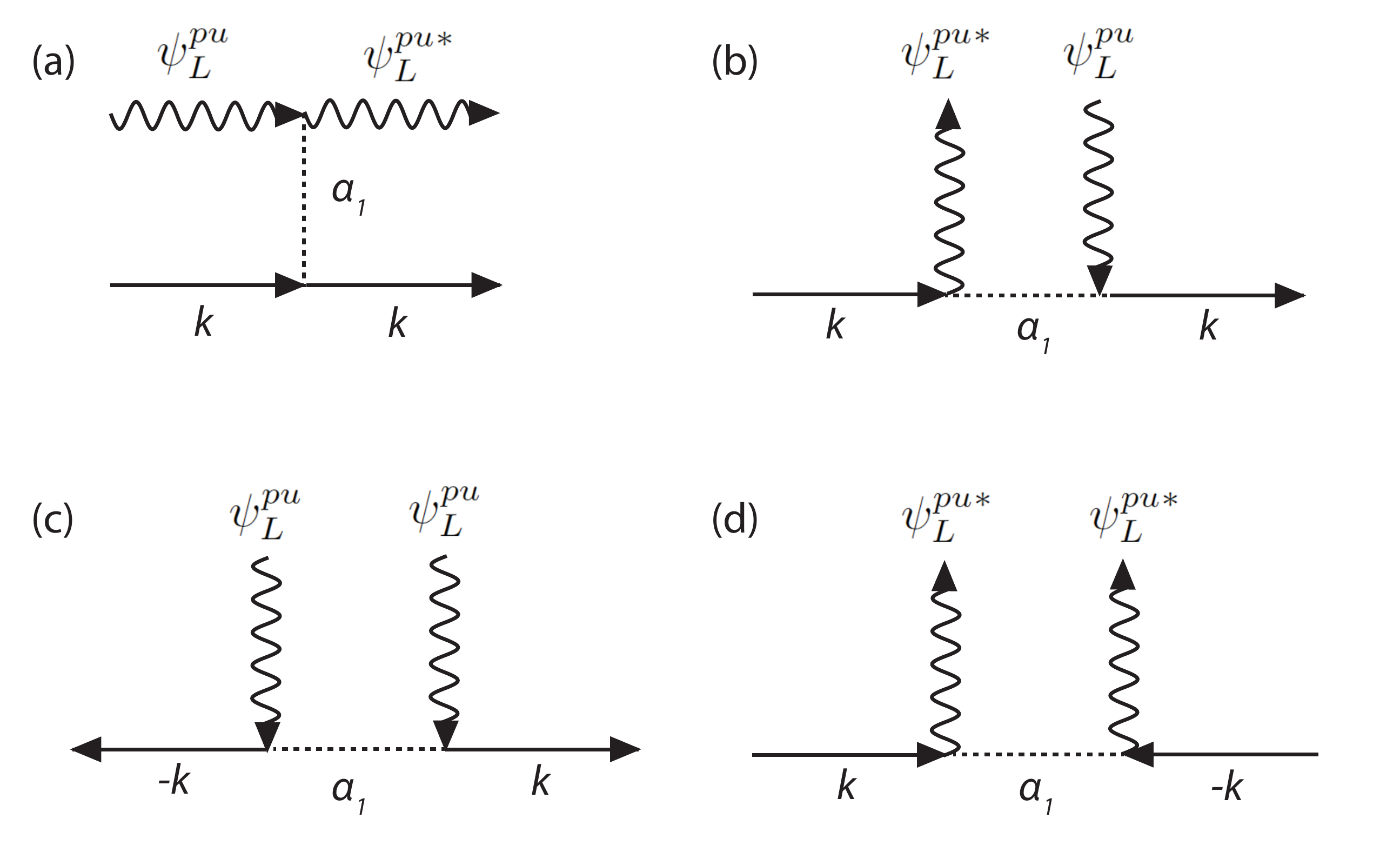}
\caption{Diagrammatic representations of the polariton interaction with parallel spins. Solid lines represent probe polaritons with momentum ${\bm k}$ (right arrow) and ${-\bm k}$ (left arrow). The wavy lines represent condensates of the lower polaritons with a zero momentum, which is driven by the strong pump beam. The dashed lines represent the polariton-polariton interaction with parallel spins $\alpha_1$. (a) Hartree-type (direct) process. (b) Fock-type (exchange) process. (c,d) The processes involving the mixing between ${\bm k}$ and ${-\bm k}$ lower polaritons (Four-wave mixing).}
\label{fig:co_diagram}
\end{figure}
\\

\section{Experimental results and analysis}
\subsection{Pump-probe spectroscopy with cocircularly polarized light}

Let us first consider the case where the pump and the
probe beams are cocircularly polarized. In this configuration, the biexciton cannot be generated; thus, we omit the spin index terms and assume that the biexciton  probability amplitude is zero. Now, the equations of motion Eq. \ref{GP_LPcr} and \ref{GP_BX} are reduced to the conventional non-equilibrium Gross-Pitaevskii equation for polaritons \cite{Carusotto2004}:  
\begin{eqnarray}
i\hbar\dot{\psi}_L({\bf x},t)&=&\Bigl[\varepsilon_L-i\gamma_L-\frac{\hbar^2\nabla^2}{2m_L}+\alpha_1|\psi_L({\bf x},t)|^2\Bigr]\psi_L({\bf x},t)\nonumber\\
& &+f_{ext}({\bf x},t).
\label{GP_LP}
\end{eqnarray}
Now, we try to derive a probe spectrum of interacting
polaritons driven by a continuous-wave (CW) pump.  First, we assume that a ``condensate" exists at ${\bm k}_{pump}={\bm 0}$ which is driven by the strong pump beam. Secondly we treat the probe beam with a finite momentum ${\bm k}$ as a small perturbation:
\begin{eqnarray}
f^{pr}_{ext}({\bf x},t)=|F^{pr}|e^{-i(\tilde{\epsilon}+\varepsilon_{pu})t/\hbar}e^{i{\bm k}{\bf x}}.
\label{eq:fpr}
\end{eqnarray}
Due to the existence of the ``condensate", we can apply the well-known Bogoliubov theory for weakly-interacting bosons \cite{Carusotto2004,pitaevskii2003bose}. Following the frame work of the Bogoliubov theory, we assume the wave function to have the following form:
\begin{eqnarray}
\psi_L({\bf x},t)&=&\left[\psi^{pu}_L+u(\tilde{\epsilon})e^{-i(\tilde{\epsilon}/{\hbar})t}e^{i{\bm k}{\bf x}}+v^*(\tilde{\epsilon})e^{i(\tilde{\epsilon}/{\hbar})t}e^{-i{\bm k}{\bf x}}\right]\nonumber\\
& &\times e^{-i(\varepsilon_{pu}/\hbar)t},
\label{bog1}
\end{eqnarray}
where $u$ and $v^*$ are the counterpropagating modes coupled by the polariton-polariton interaction $\alpha_1$, and $\psi^{pu}_L$ is the ``condensate" at ${\bm k}=0$. Substituting Eq. \ref{eq:fpr} and \ref{bog1} into the Gross-Pitaevskii equation Eq. \ref{GP_LP} and keeping only terms that involve the condensate (Bogoliubov approximation), we obtain the following equations for $u(\tilde{\epsilon})$ and $v(\tilde{\epsilon})$:
\begin{widetext}  
\begin{eqnarray}
\left(\begin{array}{cc}
\tilde{\epsilon}-(\varepsilon_L-\varepsilon_{pu}+\frac{\hbar^2{\bm k}^2}{2m_L}+2\alpha_1|\psi_L^{pu}|^2)+i\gamma_L & -\alpha_1\psi_L^{{pu}2}\\
\alpha_1\psi_L^{{pu}*2} & \tilde{\epsilon}+(\varepsilon_L-\varepsilon_{pu}+\frac{\hbar^2{\bm k}^2}{2m_L}+2\alpha_1|\psi_L^{pu}|^2)-i\gamma_L\\
\end{array}\right)
\left(\begin{array}{c}
u(\tilde{\epsilon})\\
v(\tilde{\epsilon})\\
\end{array}\right)
=
\left(\begin{array}{c}
|F^{pr}|\\
0\\
\end{array}\right)
\label{Bog_mat}
\end{eqnarray}
\end{widetext}  
The factor two of the $2\alpha_1|\psi_L^{pu}|^2$ in the diagonal element of the matrix is due to the Hartree-type (direct) and the Fock-type (exchange) contributions, which are diagrammatically represented in Fig. \ref{fig:co_diagram} (a) and (b) \cite{Lifshitz1980,Fetter2003}. On the other hand, the off-diagonal matrix elements $\alpha_1\psi_L^{{pu}2}$ and $-\alpha_1\psi_L^{{pu}2*}$ are the terms associated with the scattering of lower polaritons between ${\bm k}$ and ${-\bm k}$ momentum states. In a terminology of non-linear optics, these processes correspond to the four-wave mixing process, which are also diagrammatically presented in Fig. \ref{fig:co_diagram} (c) and (d) \cite{Lifshitz1980,Fetter2003}. These three processes contribute to the probe pulse transmission spectrum. The solution for $u(\tilde{\epsilon})$ and $v(\tilde{\epsilon})$ are easily obtained respectively as 
\begin{eqnarray}
u(\tilde{\epsilon})&=&\frac{\tilde{\epsilon}+(\varepsilon_L-\varepsilon_{pu}+\frac{\hbar^2{\bm k}^2}{2m_L}+2\alpha_1|\psi_L^{pu}|^2)-i\gamma_L}{D(\tilde{\epsilon})}|F^{pr}|\nonumber\\
\label{eq:ud}
\end{eqnarray} 
and
\begin{eqnarray}
v(\tilde{\epsilon})=-\frac{\alpha_1\psi_L^{{pu}*2}}{D(\tilde{\epsilon})}|F^{pr}|.
\end{eqnarray} 
Here, $D(\tilde{\epsilon})$ is the determinant of the matrix in Eq. \ref{Bog_mat}, which is given by
\begin{eqnarray}
D(\tilde{\epsilon})&=&\left[\tilde{\epsilon}-(\varepsilon_L-\varepsilon_{pu}+\frac{\hbar^2{\bm k}^2}{2m_L}+2\alpha_1|\psi_L^{pu}|^2)+i\gamma_L\right]\nonumber\\
& &\times\left[\tilde{\epsilon}+(\varepsilon_L-\varepsilon_{pu}+\frac{\hbar^2{\bm k}^2}{2m_L}+2\alpha_1|\psi_L^{pu}|^2)-i\gamma_L\right]\nonumber\\
& &+\alpha_1^2|\psi_L^{pu}|^4.
\end{eqnarray}
Remembering that the probe  $\psi_L^{pr}(\epsilon)$ and the four-wave mixing (idler) signal  $\psi_L^{id}(\epsilon)$  are the responses of the system at ${\bm k}$  and $-{\bm k}$ respectively, they are given by
\begin{eqnarray}
\psi_L^{pr}(\epsilon)=u(\epsilon-\varepsilon_{pu})e^{-i(\epsilon/\hbar)t}
\end{eqnarray}
and
\begin{eqnarray}
\psi_L^{id}(\epsilon)=v^*(\epsilon-\varepsilon_{pu})e^{-i((2\varepsilon_{pu}-\epsilon)/\hbar)t},
\end{eqnarray}
where we used $\tilde{\epsilon}=\epsilon-\varepsilon_{pu}$. Setting $\gamma_L=0$, the peak energies of the probe pulse transmission resonances are easily calculated  as 
\begin{eqnarray}
\varepsilon^{\pm}&=&\varepsilon_L\nonumber\\
& &\pm \sqrt{(\varepsilon_L-\varepsilon_{pu}+\frac{\hbar^2{\bm k}^2}{2m_L}+2\alpha_1|\psi_L^{pu}|^2)^2-\alpha_1^2|\psi_L^{pu}|^4}.\nonumber\\
\label{eq:eigen_co}
\end{eqnarray}
$\varepsilon^{+}$ and $\varepsilon^{-}$ respectively correspond to the normal branch and the so-called ``ghost branch" \cite{Savvidis2001,Kohnle2011,Kohnle2012,Wouters2009}. Due to the numerator of Eq. \ref{eq:ud}, the amplitude of the ghost branch is much weaker than that of the normal branch in the probe spectrum \cite{Wouters2009}. Experimentally, the ghost branch is only visible in the four-wave mixing signal \cite{Savvidis2001,Kohnle2011,Kohnle2012,Wouters2009,Zajac2015}. Therefore, in our pump-probe spectroscopy, let us consider only the normal branch. When the in-plane momentum of the probe beam is small enough (${\bm k}\simeq 0$) and the energy of the pump beam is resonant to the lower polariton branch ($\varepsilon_{pu}=\varepsilon_L$), from Eq. \ref{eq:eigen_co}, the normal branch energy shift of the probe spectrum takes the following expression derived from Eq. \ref{eq:eigen_co}:
\begin{equation}
\Delta E_{co}=\sqrt{3}\alpha_1|\psi_L^{pu}|^2.
\label{eq:co_shift}
\end{equation}
\begin{figure}
\includegraphics[width=0.5\textwidth]{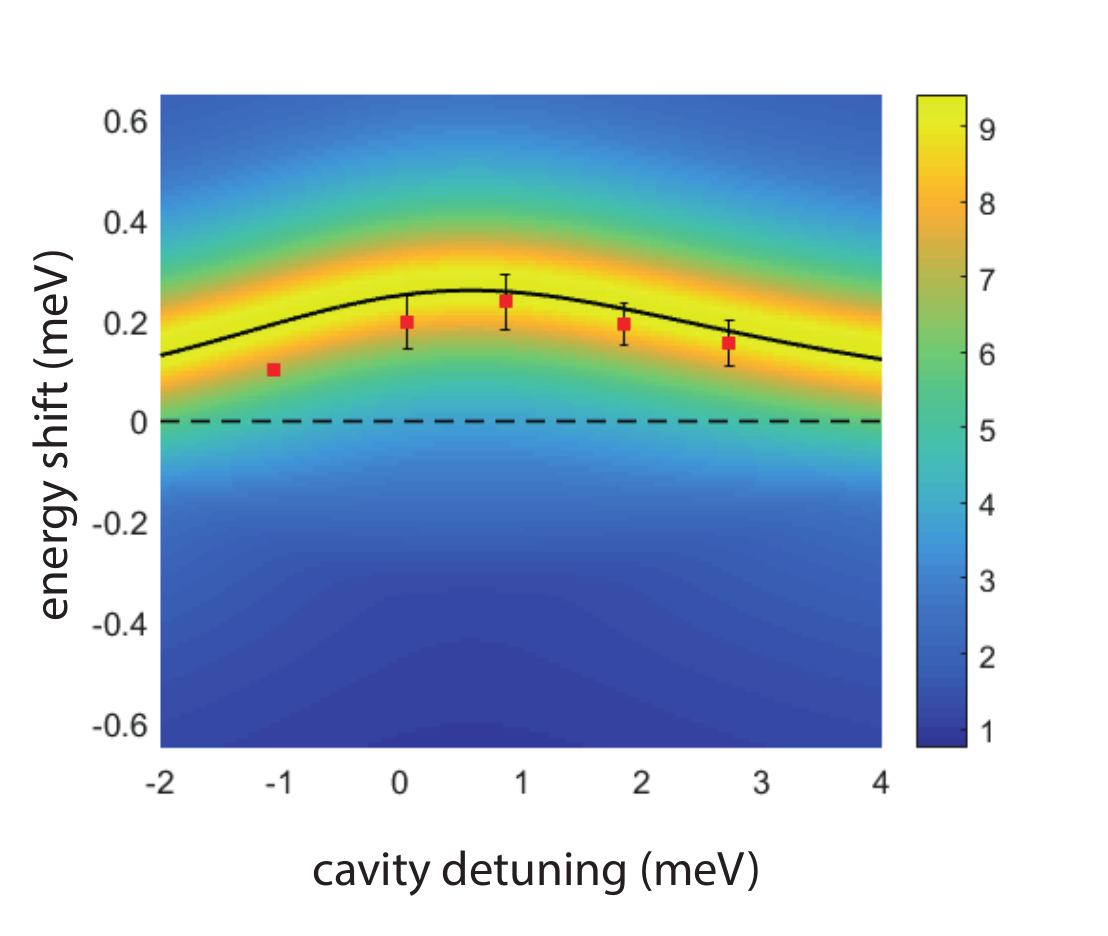}
\caption{Probe polariton peak spectrum $|\psi_{L}^{pr}(\epsilon-\varepsilon_L)|/|F^{pr}|$ is plotted as a function of the cavity detuning for a cocircularly polarized pump-probe configuration. The red filled squares are experimentally measured energy shifts of the probe peak. The solid black curve represents $\Delta E_{co}$ (Eq. \ref{eq:co_shift}). In the experiment, the pump photon mean number is set as 15$\times$10$^{12}$ photons/pulse/cm$^{2}$ (1 mW) for all detunings.}
\label{fig:co}
\end{figure}
\noindent We note that the Hartree-type, Fock-type and four-wave mixing contributions magnify the blue shift of the probe polariton by a factor of $\sqrt{3}$. The Hartree-type and Fock-type terms enhance the blue shift by a factor of two, but the population loss due to the four-wave mixing signal generation reduces the enhancement to $\sqrt{3}$. In Fig. \ref{fig:co}, the red filled squares are the experimentally extracted probe peak energy shifts at five different cavity detunings. The theoretical probe spectrum $|\psi_{L}^{pr}(\epsilon-\varepsilon_L)|/|F^{pr}|$ is also plotted as a function of the cavity detuning. Here we assume that the pump polariton population depends on the cavity detuning as $|\psi_L^{pu}|^2=|C_{\bm 0}|^2n^{pu}_{ext}$, where $n^{pu}_{ext}$ is proportional to the photon number of the pump pulse used in the experiment. As presented in Fig. \ref{fig:co}, with the increase of the cavity detuning, the energy shift becomes larger and reaches a maximum at a cavity detuning of 1 meV. This is because the lower-polariton becomes more excitonic. However, as the cavity detuning increases further, the energy shift starts to decrease, because, following $|C_{\bm 0}|^2$, the coupling between lower polaritons and the photon field outside the cavity becomes weaker .

\subsection{Pump-probe spectroscopy with countercircularly polarized lights}
\begin{figure}
\includegraphics[width=0.5\textwidth]{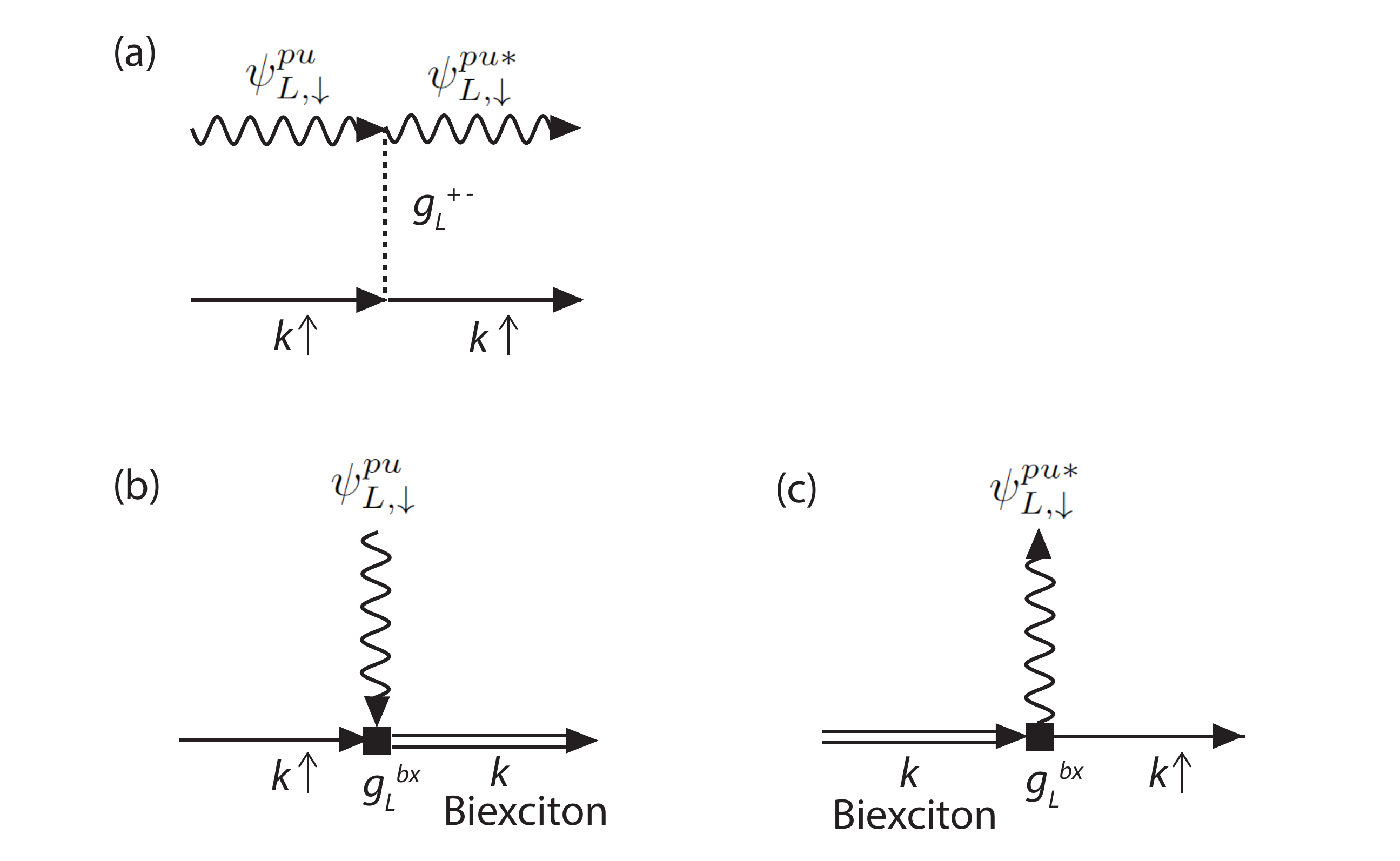}
\caption{Diagrammatic representations of the polariton interaction with anti-parallel spins. Solid lines represent spin-up probe polaritons with momentum ${\bm k}$ (right arrow), which is injected by the weak probe beam with a $\sigma^+$ circular polarization. The wavy lines represent condensates of the spin-down lower polaritons with zero momentum, which are driven by the strong pump beam with a $\sigma^-$ circular polarization. The dashed line represent the polariton-polariton interaction with anti-parallel spins $g^{\scriptscriptstyle +-}_L$. The double lines are biexcitons. (a) Hartree-type (direct) process. The association (b) and dissociation (c) of a biexciton via the polariton-biexciton coupling $g^{bx}_{L}$.}
\label{fig:cr_diagram}
\end{figure}

\begin{figure}
\includegraphics[width=0.5\textwidth]{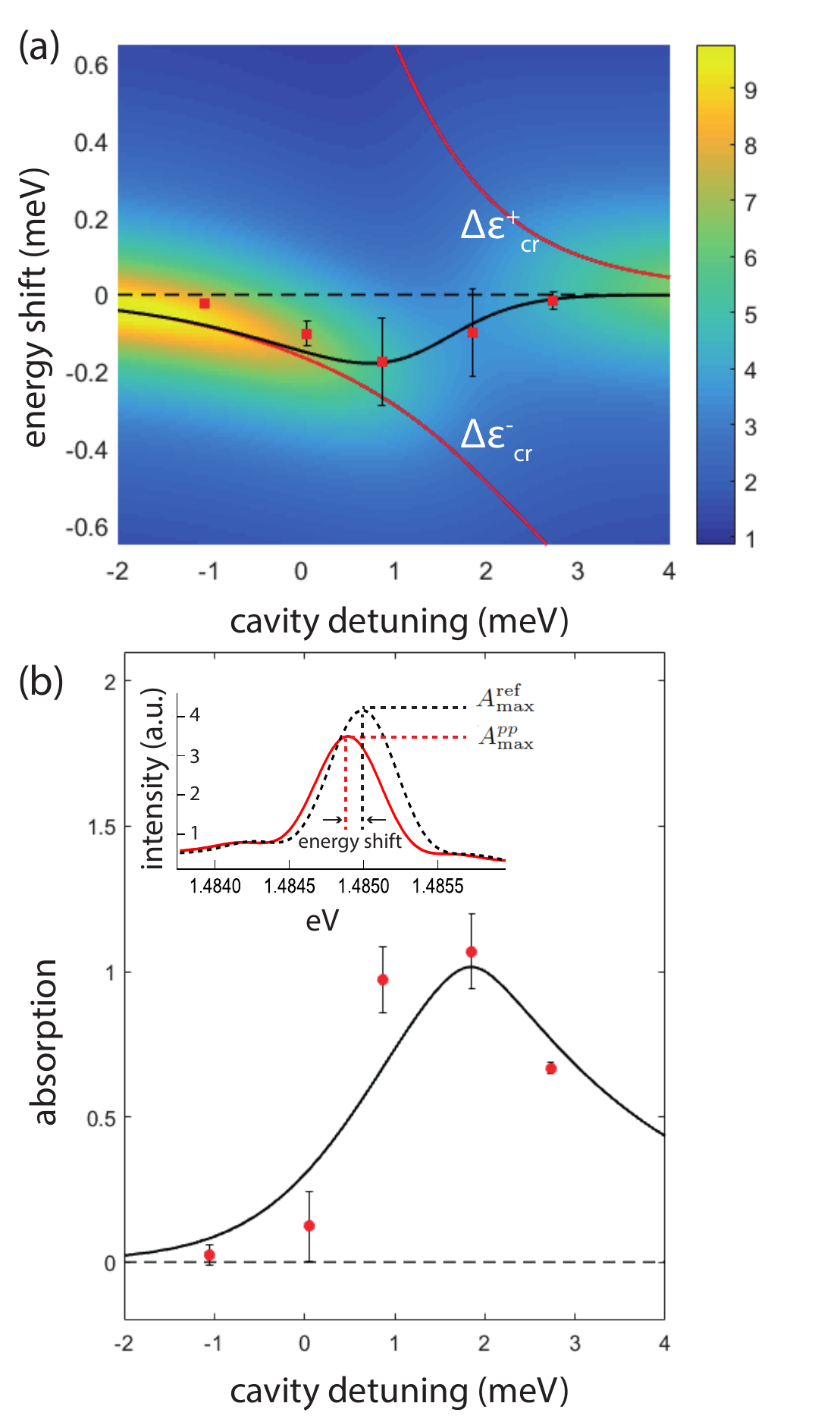}
\caption{Peak energy shift and probe pulse transmission intensity for a cross-circularly polarized pump-probe configuration. (a) The filled red squares are measured probe peak energy shifts. Probe spectrum $|\psi_{L,\uparrow}^{pr}(\epsilon-\varepsilon_L)|/|F^{pr}_\uparrow|$ is plotted as a function of the cavity detuning. The two red lines represent the energy shifts expressed as $\Delta\varepsilon_{cr}^{\pm}$ (Eq. \ref{eq:eigen_shift_cr}), while the black line is the approximated energy shift $\Delta E_{cr}$ (Eq. \ref{eq:eigen_cr}). (b) The filled red squares are measured peak absorbance. The solid black line represents the simulated absorbance based on the theoretical model. The inset presents the experimental probe spectrum with (red line) and without the pump beam (black dashed line) at the cavity detuning $\delta=0.05$ meV. In the experiment, the pump photon number is set as 15$\times$10$^{12}$ photons/pulse/cm$^{2}$ (1 mW) for all detunings.}
\label{fig:cr}
\end{figure}

Now, let us consider the case when the pump and probe pulses are countercircularly polarized. Since in this configuration the biexciton formation is allowed, we take into account the biexciton wave function and the polariton-biexciton coupling. In this section, we investigate the scattering resonance of polaritons with a biexciton within a mean-field two-channel model \citep{Takemura2014,Takemura2014a,Casteels2014}. We consider the case where the system is driven by a strong spin-down pump and probed by a weak spin-up probe. Let us assume spin-up and spin-down polariton wave-functions respectively as
\begin{eqnarray}
\psi_{L,\uparrow}=\psi^{pr}_{L,\uparrow}=u_{\uparrow}(\tilde{\epsilon})e^{-i(\tilde{\epsilon}+\varepsilon_{pu})t/\hbar}e^{i{\bm k}{\bf x}}
\end{eqnarray}
and
\begin{eqnarray}
\psi_{L,\downarrow}=\psi^{pu}_{L,\downarrow}e^{-i\varepsilon_{pu}t/\hbar}.
\end{eqnarray}
Here, the biexciton wave-function is assumed to be
\begin{eqnarray}
\psi_{B}=m(\tilde{\epsilon})e^{-i(\tilde{\epsilon}+2\varepsilon_{pu})t/\hbar}e^{i{\bm k}{\bf x}}.
\end{eqnarray}
The input probe beam is set as
\begin{eqnarray}
f^{pr}_{ext}=|F^{pr}_\uparrow|e^{-i(\tilde{\epsilon}+\varepsilon_{pu})t/\hbar}e^{i{\bm k}{\bf x}}.
\end{eqnarray}
Substituting these into the spinor Gross-Pitaevskii equations Eq. \ref{GP_LPcr} and \ref{GP_BX}, and neglecting a self mean field energy shift of the probe polaritons $\alpha_1|\psi^{pr}_{L,\uparrow}|^2$, we obtain equations for $u_\uparrow(\tilde{\epsilon})$ and $m(\tilde{\epsilon})$ as
\begin{widetext}
\begin{eqnarray}
\left(\begin{array}{cc}
\tilde{\epsilon}-(\varepsilon_L-\varepsilon_{pu}+\frac{\hbar^2{\bm k}^2}{2m_L}+g^{\scriptscriptstyle +-}_L|\psi_{L,\downarrow}^{pu}|^2)+i\gamma_L & -g^{bx}_{L}\psi_{L,\downarrow}^{pu*}\\
-g^{bx}_{L}\psi_{L,\downarrow}^{pu} & \tilde{\epsilon}-(\varepsilon_B-2\varepsilon_{pu}+\frac{\hbar^2{\bm k}^2}{2m_B})+i\gamma_B\\
\end{array}\right)
\left(\begin{array}{c}
u_\uparrow(\tilde{\epsilon})\\
m(\tilde{\epsilon})\\
\end{array}\right)
=
\left(\begin{array}{c}
|F^{pr}_\uparrow|\\
0\\
\end{array}\right).
\label{cr_meanfieled}
\end{eqnarray}
\end{widetext}
The energy shift term due to the background polariton interaction with anti-parallel spins $g^{\scriptscriptstyle +-}_L|\psi_{L,\downarrow}^{pu}|^2$ is diagrammatically represented in Fig. \ref{fig:cr_diagram} (a). We note that the Fock-type (exchange) term does not exist in $g^{\scriptscriptstyle +-}$, because the spin-up and spin-down polaritons cannot exchange indistinguishably. Additionally, the polariton-biexciton coupling terms $g^{bx}_{L}\psi_{L,\downarrow}^{pu}$ and $g^{bx}_{L}\psi_{L,\downarrow}^{pu*}$ are shown by the diagrams Fig. \ref{fig:cr_diagram} (b) and (c) respectively. Using the determinant of the matrix in Eq. \ref{cr_meanfieled}: 
\begin{eqnarray}
D^{\scriptscriptstyle +-}(\tilde{\epsilon})&=&\left[\tilde{\epsilon}-(\varepsilon_L-\varepsilon_{pu}+\frac{\hbar^2{\bm k}^2}{2m_L}+g^{\scriptscriptstyle +-}_L|\psi_{L,\downarrow}^{pu}|^2)+i\gamma_L\right]\nonumber\\
& &\times\left[\tilde{\epsilon}-(\varepsilon_B-2\varepsilon_{pu}+\frac{\hbar^2{\bm k}^2}{2m_B})+i\gamma_B\right]\nonumber\\
& &-g^{bx 2}_{L}|\psi_{L,\downarrow}^{pu}|^2,
\end{eqnarray}
the solution for $u_\uparrow(\tilde{\epsilon})$ is easily found as
\begin{eqnarray}
u_\uparrow(\tilde{\epsilon})&=&\frac{\tilde{\epsilon}-(\varepsilon_B-2\varepsilon_{pu}+\frac{\hbar^2{\bm k}^2}{2m_B})+i\gamma_B}{D^{\scriptscriptstyle +-}(\tilde{\epsilon})}|F^{pr}_\uparrow|.
\end{eqnarray} 
For $m(\tilde{\epsilon})$, the solution is
\begin{eqnarray}
m(\tilde{\epsilon})&=&\frac{g^{bx}_{L}\psi_{L,\downarrow}^{pu}}{D^{\scriptscriptstyle +-}(\tilde{\epsilon})}|F^{pr}_\uparrow|.
\end{eqnarray}
Using $\tilde{\epsilon}=\epsilon-\varepsilon_{pu}$, the probe polariton wave-function is represented as
\begin{eqnarray}
\psi_{L,\uparrow}^{pr}(\epsilon)&=&u_\uparrow(\epsilon-\varepsilon_{pu})e^{-i(\epsilon/\hbar)t}\nonumber\\
&=&{|F^{pr}_\uparrow|e^{-i(\epsilon/\hbar)t}}\nonumber\\
& &\times\left[\epsilon-(\varepsilon_L+\frac{\hbar^2{\bm k}^2}{2m_L}+g^{\scriptscriptstyle +-}_L|\psi_{L,\downarrow}^{pu}|^2)+i\gamma_L\right.\nonumber\\
& &\left.-{\displaystyle \frac{g^{bx 2}_{L}|\psi_{L,\downarrow}^{pu}|^2}{\epsilon-(\varepsilon_B-\varepsilon_{pu}+\frac{\hbar^2{\bm k}^2}{2m_B})+i\gamma_B}}\right]^{-1}
\label{cr_spectrum}
\end{eqnarray}
The above expression indicates that a resonance occurs when the sum of the energies of the pump and the probe lower polariton ($\varepsilon_{pu}+\epsilon$) coincides with that of the biexciton ($\varepsilon_B$); this is interpreted as a polaritonic version of the Feshbach resonance in cold atoms \cite{Wouters2007,Takemura2014,Casteels2014}.  Furthermore, there are two solutions corresponding to the two eigenenergies of the probe polariton states $\psi_{L,\uparrow}^{pr}$. Setting $\gamma_L=0$ and $\gamma_B=0$, the two peak energies are given by
\begin{eqnarray}
\varepsilon_{cr}^{\pm}&=&\frac{1}{2}(\varepsilon_L+\frac{\hbar^2{\bm k}^2}{2m_L}+\varepsilon_B+\frac{\hbar^2{\bm k}^2}{2m_B}-\varepsilon_{pu}+g^{\scriptscriptstyle +-}_L|\psi_{L,\downarrow}^{pu}|^2)\nonumber\\
& &\pm\frac{1}{2}\Biggl[(\varepsilon_L+\frac{\hbar^2{\bm k}^2}{2m_L}-\varepsilon_B-\frac{\hbar^2{\bm k}^2}{2m_B}+\varepsilon_{pu}\nonumber\\
& &+g^{\scriptscriptstyle +-}_L|\psi_{L,\downarrow}^{pu}|^2)^2+4g_L^{bx 2}|\psi_{L,\downarrow}^{pu}|^2\Biggr]^{\frac{1}{2}}.
\label{eq:creigen}
\end{eqnarray}
Figure. \ref{fig:cr} (a) presents the experimental energy shifts of the probe peak as red filled squares at five different values of the cavity detuning. The spectrum of the probe polariton $|\psi_{L,\uparrow}^{pr}(\epsilon-\varepsilon_L)|/|F^{pr}_\uparrow|$ is also overlapped in Fig. \ref{fig:cr} (a) as a function of the cavity detuning. The pump beam photon number $n^{pu}_{ext}$ is identical to the cocircular polarization configuration (Fig. \ref{fig:co}) and we assume $|\psi_{L,\downarrow}^{pu}|^2=|C_{\bm 0}|^2n^{pu}_{ext}$.

Supposing a small in-plane momentum of the probe polariton ${\bm k}=0$ and the pump beam resonant to the lower polariton $\varepsilon_{pu}=\varepsilon_L$, from Eq. \ref{eq:creigen}, we obtain energy shifts of the probe as 
\begin{eqnarray}
\Delta\varepsilon_{cr}^{\pm}&=&-\varepsilon_L+\frac{1}{2}(g^{\scriptscriptstyle +-}_L|\psi_{L,\downarrow}^{pu}|^2+\varepsilon_B)\nonumber\\
& &\pm\frac{1}{2}\sqrt{(g^{\scriptscriptstyle +-}_L|\psi_{L,\downarrow}^{pu}|^2-\varepsilon_B+2\varepsilon_L)^2+4g_L^{bx 2}|\psi_{L,\downarrow}^{pu}|^2}.\nonumber\\
\label{eq:eigen_shift_cr}
\end{eqnarray}
When the dephasing rate of the biexciton is large, which is the case in our system, we approximate the two modes eigenstates as a single solution by putting $\epsilon=\varepsilon^{pu}=\varepsilon_L$ into the denominator of the resonance term in Eq. \ref{cr_spectrum}. Then, the energy shift of the lower polariton mode at ${\bm k}=0$ is approximated as
\begin{equation}
\Delta E_{cr}\simeq g^{\scriptscriptstyle +-}_L|\psi_{L,\downarrow}^{pu}|^2+{\displaystyle {\rm Re}\left[\frac{g^{bx 2}_{L}|\psi_{L,\downarrow}^{pu}|^2}{2\varepsilon_{L}-\varepsilon_B+i\gamma_B}\right]}.
\label{eq:eigen_cr}
\end{equation}
Now, let us define a complex interaction constant $\alpha_2-i\alpha'_2$ based on Eq. \ref{eq:eigen_cr}. The real and imaginary parts are respectively defined as
\begin{equation}
\alpha_2=g^{\scriptscriptstyle +-}_L+g^{bx 2}_{L}\frac{2\varepsilon_{L}-\varepsilon_B}{(2\varepsilon_{L}-\varepsilon_B)^2+\gamma_B^2}
\label{eq:alpha2}
\end{equation}
and
\begin{equation}
\alpha'_2=g^{bx 2}_{L}\frac{\gamma_B}{(2\varepsilon_{L}-\varepsilon_B)^2+\gamma_B^2}.
\end{equation}
\begin{figure}
\includegraphics[width=0.5\textwidth]{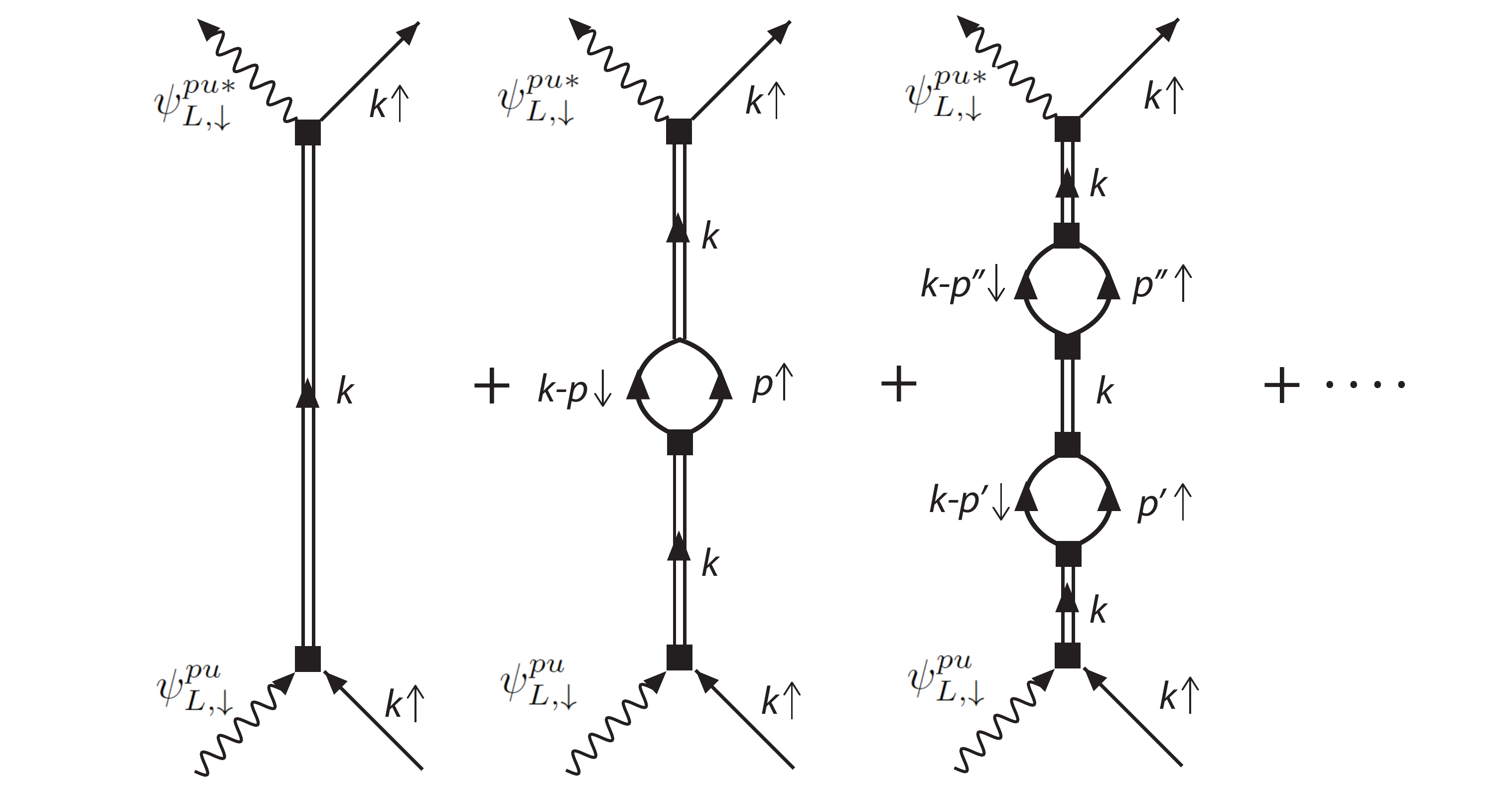}
\caption{Diagrammatic representation of the decay of a biexciton $\gamma_B$ through a spontaneous dissociation. The decay of biexciton $\gamma_B$ appears through ``renormalization''. The arrow and the double line respectively represent a lower polariton and a biexciton. The wavy lines represent condensates of the lower polaritons with a zero momentum}
\label{fig:radiative_diagram}
\end{figure}
While the energy shift of the probe spectrum is associated with  $\alpha_2$, the imaginary part of the effective interaction $\alpha'_2$ contributes to the pump induced suppression (absorption) of the spectral peak through the large biexciton decay rate $\gamma_B$. Now using the effective interaction $\alpha_2$ and $\alpha_1$, we can approximate Eq. \ref{GP_LPcr} and \ref{GP_BX} as a conventional spinor Gross-Pitaevskii equation\cite{Shelykh2006}: 
\begin{eqnarray}
i\hbar\dot{\psi}_{L,\sigma}({\bf x},t)&=&\Bigl[\varepsilon_L-i\gamma_L-\frac{\hbar^2\nabla^2}{2m_L}+\alpha_1|\psi_{L,\sigma}({\bf x},t)|^2\nonumber\\
& &+(\alpha_2-i\alpha'_2)|\psi_{L,-\sigma}({\bf x},t)|^2\Bigr]\psi_{L,\sigma}({\bf x},t)\nonumber\\
& &+f_{ext,\sigma}({\bf x},t),
\end{eqnarray}
which does not include the biexciton wave function explicitly. In Appendix. C, we discuss that this effective interaction constant $\alpha_2$ is directly obtained from the polariton-biexciton coupling Hamiltonian using a canonical transformation and interpreted as a second-order perturbation. The reduction of the spectral peak amplitude is represented as an absorbance in Fig. \ref{fig:cr} (b), which measures $ln(A_{\rm max}^{\rm ref}/A_{\rm max}^{pp})$. $A_{\rm max}^{\rm ref}$ and $A_{\rm max}^{pp}$ are maximum heights of the measured probe spectrum without and with the pump beam respectively (See the inset in Fig. \ref{fig:cr} (b)). The solid curve in Fig. \ref{fig:cr} (b) is the theoretical prediction based on the solution given by Eq. \ref{cr_spectrum}. Figure. \ref{fig:cr} (b) evidences a clear maximum of the absorption for a detuning around 2 meV, which is the consequence of the decay of polaritons through the biexciton channel. The polaritonic Feshbach resonance might be thought of the analog of the optical Feshbach resonance in the cold atom physics. In the conventional magnetic field induced Feshbach resonance, the lifetime of molecules is very long and the interaction strength diverges. In contrast, in an optical Feshbach resonance, a large loss of atoms is observed through the molecular decay channel and prevents the interaction strength from diverging \cite{Theis2004}. The large $\gamma_B$ might be associated with the spontaneous dissociation channel of the biexciton into two polaritons to all momentum combinations satisfying the energy-momentum conservation, which is represented in Fig. \ref{fig:radiative_diagram}\cite{Cohen-Tannoudji1992,Cohen-Tannoudji2011}. We may refer to this decay process as ``radiative decay'' \cite{Ivanov1995a,Ivanov2004}. The bare biexciton energy $\varepsilon_B$ is replaced with $\varepsilon_B-i\gamma_B$ through ``renormalization''\cite{Cohen-Tannoudji1992,Cohen-Tannoudji2011}. Finally, we would like to comment on the necessity of the background attractive interaction $g^{\scriptscriptstyle +-}_L$. Figure. \ref{fig:cr} (a) presents the offset of the dispersive curve, which is explained phenomenologically  with an additional background interaction. We discuss the physical origin of this background interaction in Appendix B.   
\begin{figure}
\includegraphics[width=0.5\textwidth]{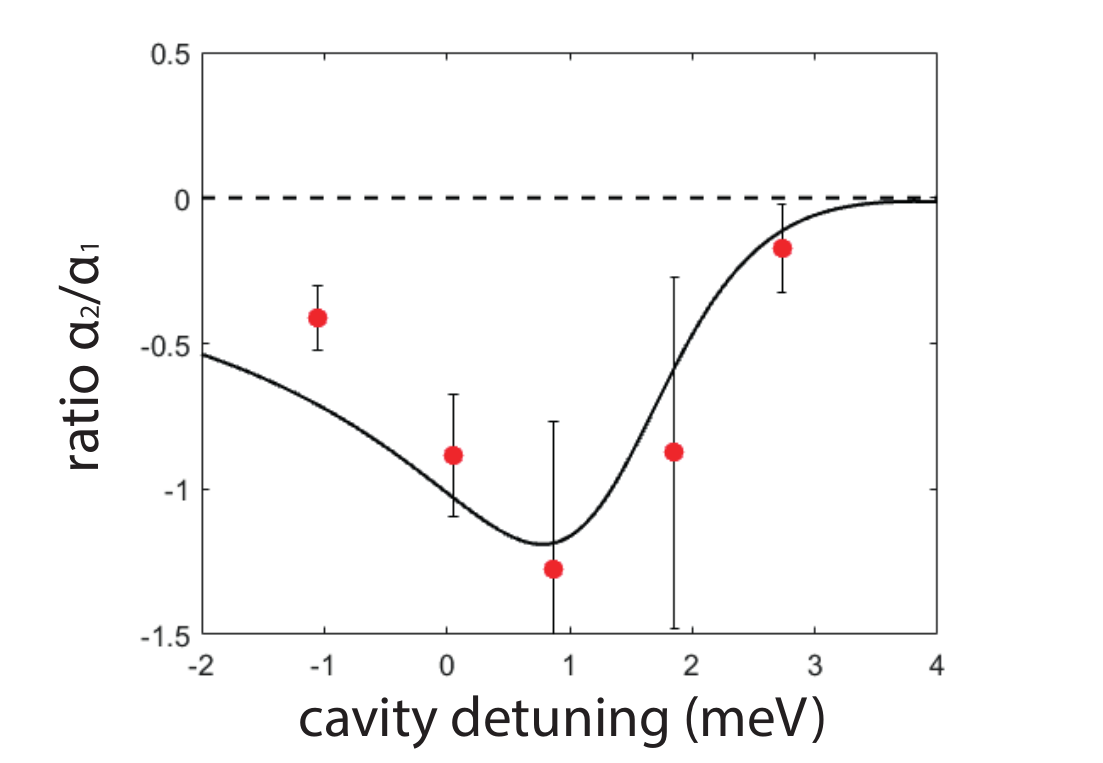}
\caption{Ratio of the effective interaction constants of lower polaritons with anti-parallel and parallel spins $\alpha_2/\alpha_1$ as a function of the cavity detuning. The solid line is the plot based on the fitting with the theoretical model.}
\label{fig:ratio}
\end{figure}

\subsection{Extraction of interaction constants and spinor interaction ratio $\alpha_2/\alpha_1$}
The physical parameters employed in the numerical plots of Fig. \ref{fig:co} and \ref{fig:cr} are extracted by fitting the model to the experiments. The fitting is performed in an automatic way by applying the least-square method to the energy shifts of cocirculary and countercircular polarization configuration, the absorbance, and the ratio $\alpha_2/\alpha_1$ simultaneously. For the fitting, we fix the exciton-exciton interaction constant $g$ as a unity ($g\equiv 1$), then the pump photon density is found to be $n^{pu}_{\rm ext}=0.53$. The extracted interaction constants are scaled with respect to the value of g: i.e., the polariton interaction constants $g^{\scriptscriptstyle +-}=-1.19g$ and $g^{bx}=1.37g\sqrt{n^{pu}_{\rm ext}}$. The extracted biexciton energy and dephasing rate are respectively $\varepsilon_B=2\varepsilon_x-2.13$ meV and $\gamma_B=0.80$ meV. On the other hand, for simplicity, we fix the values of photon assisted exchange scattering strength and lower polariton dephasing rate to $g_{\rm pae}=0.3g$ and $\gamma_L=0.1$ meV respectively \cite{Takemura2015b,Ciuti2000,Brichkin2011}.  The parameters used for the theoretical plots are summarized in Table. \ref{table:parameters}
\begin{table}[hbtp]
  \caption{Parameters used for the theoretical curves. The binding energy of a biexciton $\varepsilon_{\rm bin}$ is defined as $\varepsilon_{\rm bin}=2\varepsilon_x-\varepsilon_B$. The parameters $g$, $g_{\rm pae}$, and $\gamma_L$ have fixed values.}
  \label{table:data_type}
  \centering
  \begin{tabular}{cccccccc}
  \\
    \hline \hline
    $g$  & $g_{\rm pae}$  & $g^{\scriptscriptstyle +-}$ & $g^{bx}$ & $\gamma_L$ & $\gamma_B$ & $\varepsilon_{\rm bin}$ & $n^{pu}_{\rm ext}$\\
    & ($g$)  & ($g$) & ($g\sqrt{n^{pu}_{\rm ext}}$) & (meV) & (meV) & (meV) & \\
    \hline 
    1 & 0.3 & -1.19 & 1.37 & 0.1 & 0.80 & 2.13 & 0.53\\
    \hline \hline
  \end{tabular}
\label{table:parameters}  
\end{table} 

Finally, we present the ratio between polariton interaction with parallel and with anti-parallel spins. Using the effective interaction constant $\alpha_2$ defined in Eq. \ref{eq:alpha2} and $\alpha_1$ in Eq. \ref{eq:alpha1}, the ratio of the polariton interaction $\alpha_2/\alpha_1$ is shown in Fig. \ref{fig:ratio}. Experimentally, the ratio of the interactions is obtained from the energy shifts as $\alpha_2/\alpha_1=\sqrt{3}\Delta E_{cr}^{\rm expt}/\Delta E_{co}^{\rm expt}$, where $\Delta E_{co(cr)}^{\rm expt}$ is the measured probe energy shift in the co (counter)-circularly polarized pump-probe configuration presented in Fig. \ref{fig:co} (Fig. \ref{fig:cr}). The factor $\sqrt{3}$ is necessary because of an enhancement of the probe energy shift $\Delta E_{co}^{\rm expt}$ (Eq. \ref{eq:co_shift}). An important consequence of the cavity detuning dependence of the ratio $\alpha_2/\alpha_1$ is that the lower polariton interaction with anti-parallel spins enhances dramatically and becomes comparable to that of parallel spins in the vicinity of the biexciton resonance \cite{Vladimirova2010,Takemura2014a}. This indicates that the collapse of Bose-Einstein condensates of polaritons might be possible at positive cavity detuning \cite{Vladimirova2010}. However, we stress that in the proximity of the scattering resonance the enhanced non-linear decay is not negligible, which is associated with the imaginary part of the interaction constant $\alpha_2'$. Even though the enhancement of $\alpha_2$ is not as dramatic as the Feshbach resonance in cold atoms, the large dissipative non-linearity $\alpha_2'$ is still useful in the non-equilibrium system for some applications such as non-classical photon generation \cite{Carusotto2010} and polariton spin switching \cite{Amo2010}. Actually, the use of the dissipative non-linearity was pointed out in the context of a polariton blockade \cite{Carusotto2010}. Another interesting direction of investigation in relation with the large biexciton decay rate is a source of entangled photons. As discussed in the previous subsection, the large decay rate of the biexciton might originate from the spontaneous dissociation channel into two polaritons (radiative decay). If the decay rate of the biexciton $\gamma_B$ is mainly due to the radiative decay, as predicted in Ref \cite{Oka2008}, the biexciton will produce entangled polaritons (or photons outside the microcavity) with anti-parallel spins (with countercircular polarizations) in its radiative decay. If an additional decay process exists in $\gamma_B$, it will degrade the efficiency of the entangled photon generation.    

\section{Conclusion}
We investigated spinor interactions of lower polaritons in semiconductor microcavities employing a pump-probe technique with a spectrally narrow pulse. The analysis of the stationary properties of interacting lower polaritons was performed with the Bogoliubov theory and a mean-field two-channel model. We observed a large enhancement of the interaction strength of the lower polaritons with opposite spins in the vicinity of the scattering resonance with the biexciton. This phenomenon can be interpreted as a polaritonic version of the Feshbach resonance in cold atom physics. Furthermore, the ratio between the lower polariton interaction with parallel ($\alpha_1$) and anti-parallel spins ($\alpha_2$) indicates that $\alpha_2$ can be comparable to $\alpha_1$ in the proximity of the scattering resonance. 

\section*{Acknowledgments}
This work is supported by the Swiss National Science Foundation under Project No. 153620 and the European Research Council under project Polaritonics, Contract No. 291120. The polatom network is also acknowledged.
\section*{Appendix A: Spinor polariton dynamics}
\begin{figure*}
\includegraphics[width=0.85\textwidth]{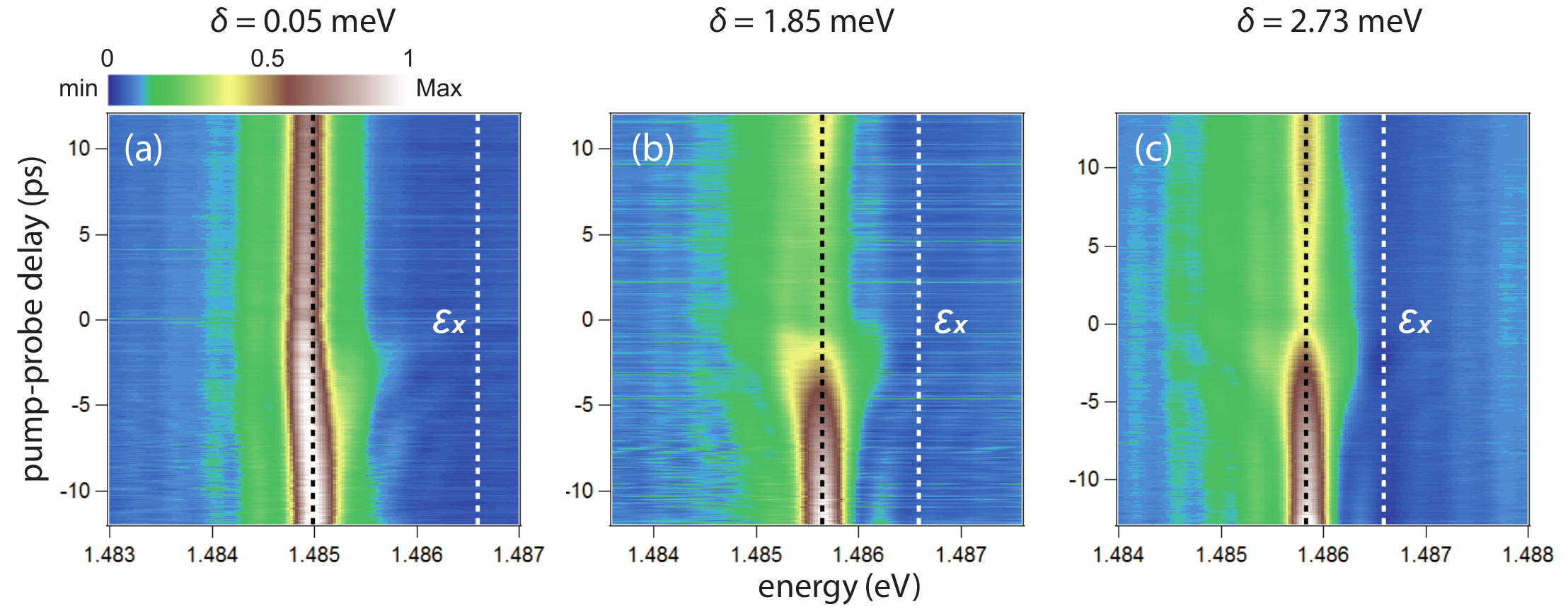}
\caption{Experimental probe spectra as a function of the pump-probe delay for the cavity detuning: 0.05 meV (a), 1.85 meV (b), and 2.73 meV (c). The black dashed lines represent the lower polariton peak energies without the pump pulse. The exciton energy is represented as the white dashed line. The pump photon number is set as 15$\times$10$^{12}$ photons/pulse/cm$^{2}$ (1 mW) for all detunings. The color scale is defined in Fig. \ref{fig:dmap_sim}}
\label{fig:dmap_exp}
\end{figure*}
\begin{figure*}
\includegraphics[width=0.85\textwidth]{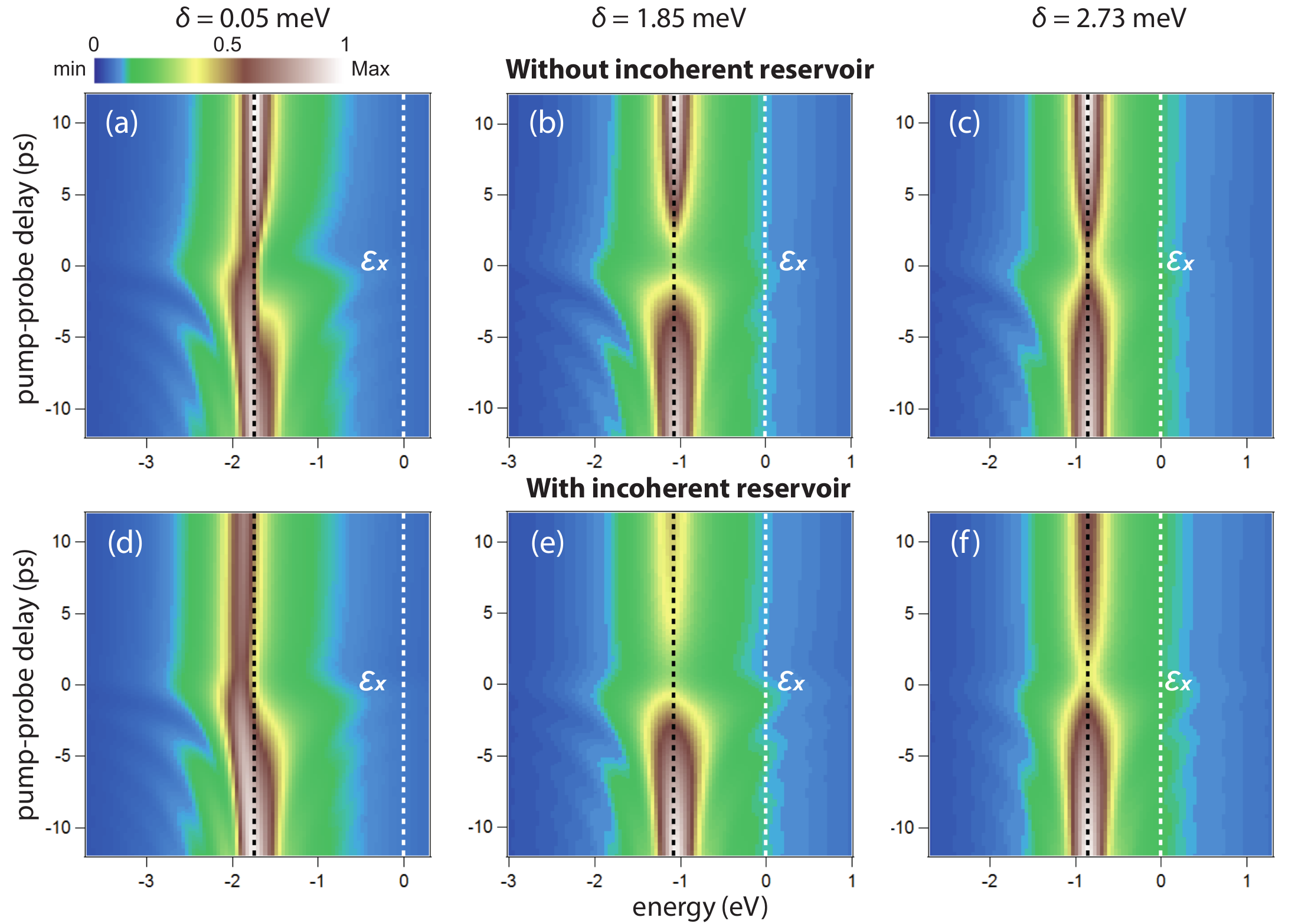}
\caption{Simulated probe spectra based on the coherent polariton-biexciton Gross-Pitaevskii equations (a-c) and the phenomenological model with an incoherent reservoir (d-f) as a function of the pump-probe time delay. The cavity detunings are 0.05 meV (a,d), 1.85 meV (b,e), and 2.73 meV (c,f). The black dashed lines represent the lower polariton peak energies without the pump pulse. The exciton energy is represented as the white dashed line. The color scale is defined by a normalization with the maximum intensity of the lower polariton branch.}
\label{fig:dmap_sim}
\end{figure*}

 
In the main text, the pump-probe time delay is fixed to zero and only the stationary properties of spinor polaritons are investigated. Here we investigate the dynamical aspects of the spinor polaritons. The dynamical properties of the polariton-biexciton coupling system will determine the real time formation of the biexcitons in the vicinity of the polaritonic Feshbach resonance. Moreover, the time delay dependence of the pump-probe spectra provides the response time of the optical non-linearity, which ultimately determines the switching speed of polaritonic devices. Figure. \ref{fig:dmap_exp} presents probe spectra as a function of the pump-probe time delay for three different cavity detunings: below (a), on (b), and above the resonance
(c) with the biexciton. In order to simulate the dynamical properties, the simplest approach is to numerically integrate the spinor Gross-Pitaevskii equations with biexcitons (Eq. \ref{GP_LPcr} and \ref{GP_BX}). Coupled mode equations for the probe, pump, and biexciton are written as
\begin{eqnarray}
i\hbar\dot{\psi}_{L,\uparrow}^{pr}(t)&=&\Big[\varepsilon_L-i\gamma_L+\alpha_1|\psi_{L,\uparrow}^{pr}(t)|^2\nonumber\\
& &+g^{\scriptscriptstyle +-}_L|\psi_{L,\downarrow}^{pu}(t)|^2\Bigr]\psi^{pr}_{L,\uparrow}(t)\nonumber\\
& &+g^{bx}_{L}\psi_{B}(t)\psi_{L,\downarrow}^{pu*}(t)+f_{ext,\uparrow}^{pr}(t),
\end{eqnarray}
\begin{eqnarray}
i\hbar\dot{\psi}_{L,\downarrow}^{pu}(t)&=&\Bigl[\varepsilon_L-i\gamma_L+\alpha_1|\psi_{L,\downarrow}^{pu}(t)|^2\nonumber\\
& &+g^{\scriptscriptstyle +-}_L|\psi_{L,\uparrow}^{pr}(t)|^2\Bigr]\psi_{L,\downarrow}^{pu}(t)\nonumber\\
& &+g^{bx}_{L}\psi_{B}(t)\psi_{L,\uparrow}^{pr*}(t)+f_{ext,\downarrow}^{pu}(t),
\end{eqnarray}
and
\begin{eqnarray}
i\hbar\dot{\psi}_{B}(t)&=&(\varepsilon_B-i\gamma_B)\psi_{B}(t)+g^{bx}_{L}\psi_{L,\uparrow}^{pr}(t)\psi_{L,\downarrow}^{pu}(t).
\end{eqnarray}
The driving pump (probe) field is written as a Gaussian pulse: 
\begin{eqnarray}
f^{pu(pr)}_{ext,\downarrow(\uparrow)}&=&F^{pu(pr)}\exp\left[-\frac{(t-t_{pu(pr)})^2}{2\tau_{pu(pr)}^2}\right]\nonumber\\
& &\cdot\exp\left[-i\frac{\varepsilon_{pu(pr)}}{\hbar}(t-t_{pu(pr)})\right].
\end{eqnarray}
We set the pulse durations as $\tau_{pu}=1.5$ ps and $\tau_{pr}$=0.35 ps. The setting of the pulse intensities are $|F^{pu}|^2=0.16$ and $|F^{pr}|^2=0.03$. Simulated probe spectra based on the above equations are shown in Fig. \ref{fig:dmap_sim} (a)-(c). All interaction constants used for the simulations in Fig. \ref{fig:dmap_sim} are the same as those used for the stationary model in the main text. As we expect, this set of equations reproduces the basic features of the polaritonic Feshbach resonance (The energy shift and absorption of the probe polariton induced by the biexciton). Nevertheless, there are striking differences between the experiments and simulations in the pump-probe delay dependence. In particular, the simulation does not reproduce the positive pump-probe delay part of the observed spectra. The cause is that the model used here is
fully coherent and neglects any long-lived population beyond the cavity lifetime. Inspired by previous works \cite{Sarkar2010,Wouters2013,Takemura2016}, we try to construct phenomenologically a set of coupled mode equations including an exciton reservoir:
\begin{eqnarray}
i\hbar\dot{\psi}_{L,\uparrow}^{pr}(t)&=&\Big[\varepsilon_L-i\gamma_L+(\alpha_{2C}-i\alpha'_{2C})\cdot|\psi_{L,\downarrow}^{pu}(t)|^2\nonumber\\
& &+(\alpha_{2R}-i\alpha'_{2R})\cdot n_{R\downarrow}(t)\Big]\psi_{L,\uparrow}^{pr}(t)+f_{ext,\uparrow}^{pr}(t),\nonumber\\
\label{GP_reservoir_up}
\end{eqnarray}
\begin{eqnarray}
i\hbar\dot{\psi}_{L,\downarrow}^{pu}(t)&=&\Big[\varepsilon_L-i\gamma_L+(\alpha_{1C}-i\alpha'_{1C})\cdot|\psi_{L,\downarrow}^{pu}(t)|^2\nonumber\\
& &+(\alpha_{1R}-i\alpha'_{1R})\cdot n_{R,\downarrow}(t)\Big]\psi_{L,\downarrow}^{pu}(t)+f_{ext,\downarrow}^{pu}(t),\nonumber\\
\label{GP_reservoir_down}
\end{eqnarray}
and
\begin{eqnarray}
\hbar\dot{n}_{R,\downarrow}(t)&=&-\gamma_Rn_{R,\downarrow}(t)+2\alpha'_{1C}|\psi_{L,\downarrow}^{pu}(t)|^4,
\label{reservoir_down}
\end{eqnarray}
\noindent where $n_{R,\downarrow}$ is the spin-down reservoir population and a long lifetime is assumed $\gamma_R=0.01$ meV \cite{Takemura2016}. Here we neglect the creation of a spin-up incoherent reservoir and a corresponding coherent population of polaritons, because the spin-up probe pulse is weak and, thus, contribute only by a small amount of spin-up excitons and polaritons to the incoherent reservoir and to the coherent population, respectively. In this approach, we do not include the biexciton degree of freedom; instead, the effect of the biexciton resonance is indirectly included in the effective complex interaction constants $\alpha_{2C(R)}$ and $\alpha'_{2C(R)}$. In addition to this, we also introduce a complex interaction constant $\alpha_{1C(R)}+i\alpha'_{1C(R)}$ for the polaritons with parallel spins. The imaginary part $\alpha'_{1C(R)}$ represents excitation induced dephasing (EID), which converts the coherent polariton population into the incoherent population \cite{Takemura2016}. The symbol $C$ and $R$ represent the contributions to the mean field energy shift from coherent polaritons and incoherent reservoir respectively. In this appendix, for simplicity, we chose the real parts as $\alpha_{1C}=\alpha_{1R}=\alpha_{1}$ and $\alpha_{2C}=\alpha_{2R}=\alpha_{2}$, and the imaginary parts as $\alpha_{1C}^{\prime}=\alpha_{1R}^{\prime}=\alpha_{1}^{\prime}$ and $\alpha_{2C}^{\prime}=\alpha_{2R}^{\prime}=\alpha_{2}^{\prime}$. Following our previous work \cite{Takemura2016}, we set $\alpha'_1=0.3g$. In this condition, the mean field energy shift is simply proportional to the total population such as $\alpha_{1(2)}N_{\downarrow}$, where the total population $N_{\downarrow}$ is given by $N_{\downarrow}=|\psi_{L,\downarrow}^{pu}|^2+n_{R\downarrow}$. We stress that this choice is just one of the possible choices of the coherent and incoherent interaction constants that can qualitatively reproduce the experimental results. Numerical simulations are presented in Fig. \ref{fig:dmap_sim} (d)-(f); they reproduce the positive pump-probe delay region of the probe spectra much better than the previous fully coherent model. In the positive pump-probe time delay part, the energy-shift and absorption of the probe polariton survives more than 10 ps. This is because the decay of the coherent spin-down
polariton population is compensated by the long-lived spin-down incoherent population in the reservoir, which is induced by EID. In other words, the long lived energy-shift and the reduction of the polariton absorption at positive pump-probe delays are associated with the coupling between the incoherent reservoir and biexciton \cite{Saba2000}. On the other hand, the negative pump-probe delay part of the spectrum mainly evidences a coherent coupling between polaritons and biexcitons \cite{Neukirch2000}. Although the incoherent reservoir model phenomenologically reproduces the experimental delay dependent pump-probe spectra, a further investigation is required in order to understand the true character and the creation mechanism of the reservoir \cite{Takemura2016}. In particular, since two lower polaritons with low in-plane momenta energetically cannot scatter to a large in-plane momentum exciton reservoir, the nature of the reservoir is not clear. For instance, one should consider the scattering process of a lower polariton from the pump pulse with an upper polariton from the probe pulse to the reservoir of excitons in order to account for the absorption change at the spectral position of the lower polariton resonance. Even if we consider this process we cannot explain the existence of a long lived reservoir. From the design point of view of polaritonic devices, the long lifetime of the reservoir ($\hbar/\gamma_R\sim 66$ ps) limits the response time; thus, the reservoir creation should be minimized by choosing appropriate experimental conditions: a low enough excitation power and a cavity detuning away from the biexciton resonance and EID.        

\section*{Appendix B: Exciton scattering continuum and back ground interaction}
In this appendix, we discuss how the background interaction appears from a bare exciton-exciton interaction. In general, using a non-contact potential $W^{\scriptscriptstyle +-}({\bf x})$ and the field operator of excitons, the bare exciton-exciton interaction is expressed as 
\begin{eqnarray}
\hat{W}^{\scriptscriptstyle +-}&=&\int d{\bf x} \left[\hat{\bm \psi}_{x,\uparrow}^{\dagger}\hat{\bm \psi}_{x,\downarrow}'^{\dagger}W^{\scriptscriptstyle +-}({\bf x}-{\bf x}')\hat{\bm \psi}'_{x,\downarrow}\hat{\bm \psi}_{x,\uparrow}\right]\nonumber\\
&=&\sum_{{\bm k}{\bm k}'{\bm q}}\tilde{W}^{\scriptscriptstyle +-}({\bm q})\hat{x}^\dagger_{{\bm k}-{\bm q},\uparrow}\hat{x}^\dagger_{{\bm k}'+{\bm q},\downarrow}\hat{x}_{{\bm k}',\downarrow}\hat{x}_{{\bm k},\uparrow}.
\end{eqnarray}
Here $\hat{x}_{\bm k,\sigma}$ is an exciton annihilation operator. The exciton annihilation operator $\hat{x}_{\bm k,\sigma}$ and the exciton field operator $\hat{\psi}_{x,\sigma}({\bf x})$ are related as
\begin{equation}
\hat{\psi}_{x,\sigma}({\bf x})=\sum_{{\bm k}} \frac{1}{\sqrt{S}}e^{{\bm k}\cdot{\bf x}}\cdot\hat{x}_{\bm k,\sigma},
\end{equation} 
where $S$ is the area occupied by excitons. $\tilde{W}^{\scriptscriptstyle +-}({\bm q})$ is defined with the Fourier transformation as
\begin{equation}
\tilde{W}^{\scriptscriptstyle +-}({\bm q})=\int W^{\scriptscriptstyle +-}({\bf x})e^{-{\bm q}\cdot{\bf x}}d{\bf x}
\end{equation} 
It is important to note that the exciton-exciton interaction matrix $\tilde{W}^{\scriptscriptstyle +-}({\bm q})$ explicitly depends on momentum, which is the consequence of the non-contact interaction. Our objective is to find an effective interaction constant which does not depend on momentum. We need to find an effective interaction which reproduces the same scattering amplitude as the bare exciton-exciton interaction. In the three dimensional case, this procedure corresponds to replacing a momentum dependent bare interaction with a momentum independent scattering length \cite{Proukakis1998,Pethick2002}. Since the scattering amplitude is connected to $T$-matrix, we evaluate the $T$-matrix using Lippman-Schwinger equation \cite{Takayama2002,Rodberg1967}
\begin{eqnarray}
\hat{T}(\epsilon)&=&\hat{W}^{\scriptscriptstyle +-}+\hat{W}^{\scriptscriptstyle +-}\frac{1}{\epsilon-\hat{H}_0}\hat{T}(\epsilon)\\
&=&\hat{W}^{\scriptscriptstyle +-}+\hat{W}^{\scriptscriptstyle +-}\frac{1}{\epsilon-\hat{H}^{\scriptscriptstyle +-}}\hat{W}^{\scriptscriptstyle +-},
\end{eqnarray}
where $\hat{H}_0$ is the kinetic term of two excitons and $\hat{H}^{\scriptscriptstyle +-}=\hat{H}_0+\hat{W}^{\scriptscriptstyle +-}$.
Let us calculate the above $T$-matrix in terms of two-exciton basis defined as 
\begin{equation}
|{\bm K} {\bm q}\rangle\equiv\hat{x}^\dagger_{{\bm K}/2+{\bm q},\uparrow}\hat{x}^\dagger_{{\bm K}/2-{\bm q},\downarrow}| 0\rangle,
\end{equation}
where the momentum ${\bm K}$ and ${\bm q}$ are respectively a center and a relative momentum given by
\begin{equation}
{\bm K}={\bm k}+{\bm p}
\end{equation}
and
\begin{equation}
{\bm q}=\frac{{\bm k}-{\bm p}}{2},
\end{equation}
where ${\bm k}$ and ${\bm p}$ are momenta of exciton with spin-up and -down, respectively. The $T$-matrix leads to    
\begin{eqnarray}
\langle {\bm K} {\bm q}_1|\hat{T}(\epsilon)|{\bm K} {\bm q}_2\rangle
&=&\langle {\bm K} {\bm q}_1|\hat{W}^{\scriptscriptstyle +-}|{\bm K} {\bm q}_2\rangle\nonumber\\
& &+\langle {\bm K} {\bm q}_1|\hat{W}^{\scriptscriptstyle +-}\frac{1}{\epsilon-\hat{H}^{\scriptscriptstyle +-}}\hat{W}^{\scriptscriptstyle +-}|{\bm K} {\bm q}_2\rangle\nonumber\\
&=&\langle {\bm K} {\bm q}_1|\hat{W}^{\scriptscriptstyle +-}|{\bm K} {\bm q}_2\rangle\nonumber\\
& &+\sum_{{\bm q}'{\bm q}''}\langle {\bm K} {\bm q}_1|\hat{W}^{\scriptscriptstyle +-}|{\bm K}{\bm q}'\rangle\nonumber\\
& &\times\langle {\bm K}{\bm q}'|\frac{1}{\epsilon-\hat{H}^{\scriptscriptstyle +-}}|{\bm K}{\bm q}''\rangle\nonumber\\
& &\times\langle {\bm K}{\bm q}''|\hat{W}^{\scriptscriptstyle +-}|{\bm K} {\bm q}_2\rangle
\label{T-matrix2}    
\end{eqnarray} 
\begin{figure}
\includegraphics[width=0.5\textwidth]{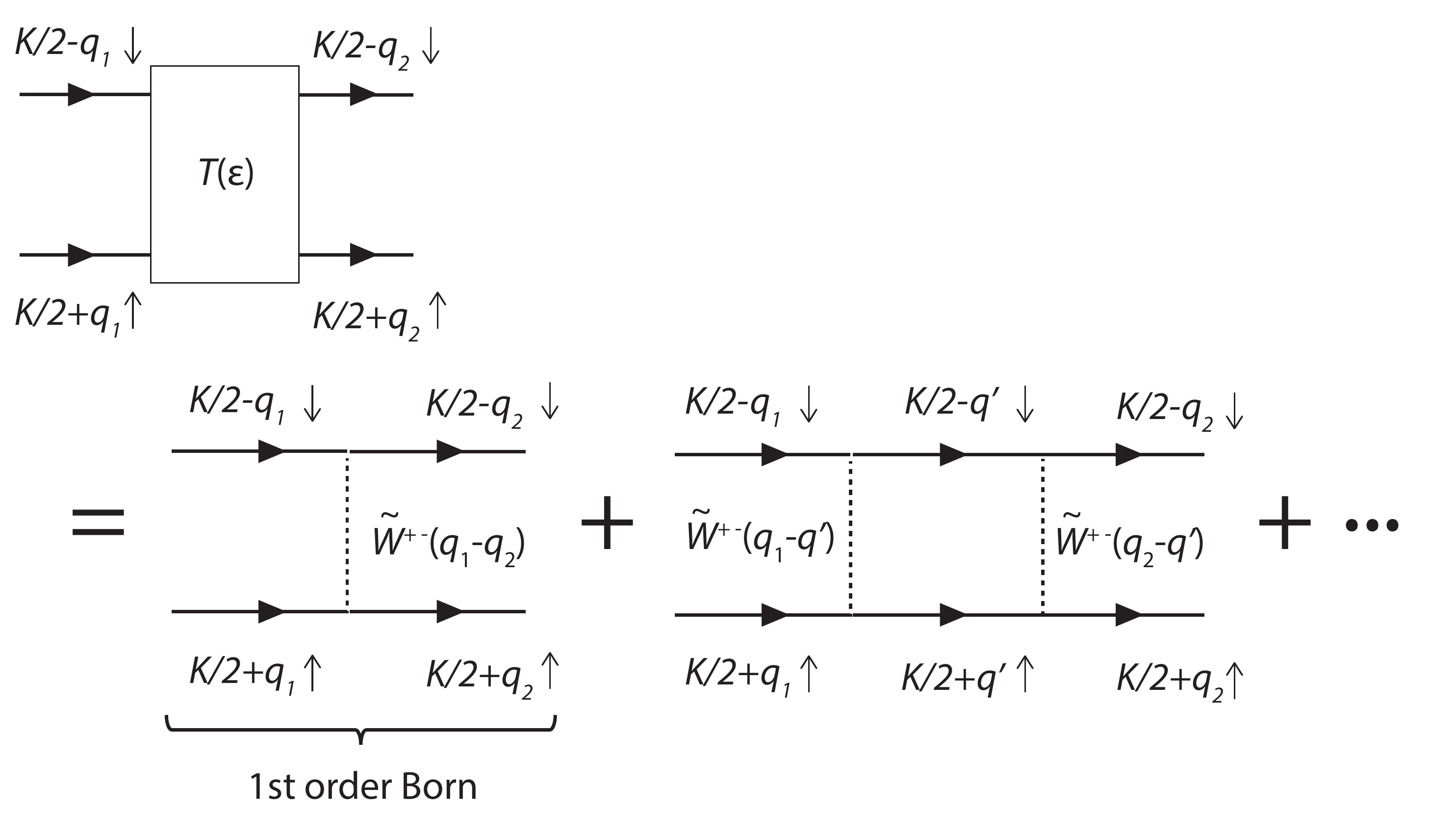}
\caption{Diagrammatic representation of the $T$-matrix.}
\label{fig:t-matrix}
\end{figure} 
The Hamiltonian $\hat{H}^{\scriptscriptstyle +-}$ is generally expanded by two-exciton states as 
\begin{eqnarray}
\frac{1}{\epsilon-\hat{H}^{\scriptscriptstyle +-}}=\sum_{n}\frac{|B^{(n)}_{\bm K}\rangle\langle B^{(n)}_{\bm K}|}{\epsilon-\Omega^{(n)}_{\bm K}}
\label{eq:two-excitons}
\end{eqnarray}
Using a two-exciton wave-function $\psi_B^{(n)}({\bm K},{\bm l})$, the nth two-exciton states $|B^{(n)}_{\bm K}\rangle$ is defined as
\begin{eqnarray}
|B^{(n)}_{\bm K}\rangle &\equiv&\sum_{{\bm l}}\psi_B^{(n)}({\bm K},{\bm l})\hat{x}^\dagger_{{\bm K}/2+{\bm l},\uparrow}\hat{x}^\dagger_{{\bm K}/2-{\bm l},\downarrow}| 0\rangle\nonumber\\
&=&\sum_{{\bm l}}\psi_B^{(n)}({\bm K},{\bm l})|{\bm K} {\bm l}\rangle
\end{eqnarray}
The $n$th two-exciton wave-function $\psi_B^{(n)}({\bm K},{\bm l})$ and eigen energy $\Omega^{(n)}_{\bm K}$ satisfy the following two-body Schroedinger equation (also referred to as the Bethe–Salpeter equation) \cite{Ivanov1995}:
\begin{eqnarray}
& &[\varepsilon_{x,\frac{1}{2}{\bm K}+{\bm l}}+\varepsilon_{x,\frac{1}{2}{\bm K}-{\bm l}}]\psi_B^{(n)}({\bm K},{\bm l})\nonumber\\
& &+\sum_{{\bm l}'}\tilde{W}^{\scriptscriptstyle +-}({\bm l}-{\bm l}')\psi_B^{(n)}({\bm K},{\bm l}')=\Omega^{(n)}_{\bm K}\psi_B^{(n)}({\bm K},{\bm l}),
\end{eqnarray}
where $\varepsilon_{x,\frac{1}{2}{\bm K}\pm{\bm l}}$ is the energy of a free exciton. This equation indicates that, if the exciton-exciton interaction potential $\tilde{W}^{\scriptscriptstyle +-}({\bm q})$ is given, both wave-functions and eigen energies are uniquely determined. Their explicit calculation is out of the scope of our paper, but we refer to several works that attempted the estimations \cite{Takayama2002,Kwong2001}. Finally, the $T$-matrix Eq. \ref{T-matrix2} is simplified as
\begin{eqnarray}
\langle {\bm K} {\bm q}_1|\hat{T}(\epsilon)|{\bm K} {\bm q}_2\rangle
&=&\tilde{W}^{\scriptscriptstyle +-}({\bm q}_1-{\bm q}_2)\nonumber\\
& &+\sum_{n}
\frac{g_{bx}^{(n)}({\bm K},{\bm q}_1,{\bm q}_2)}{\epsilon-\Omega_{\bm K}^{(n)}},
\end{eqnarray}
where $g_{bx}^{(n)}$ represents \cite{Ivanov1995}
\begin{eqnarray}
g_{bx}^{(n)}({\bm K},{\bm q}_1,{\bm q}_2)&=&\sum_{{\bm q}'{\bm q}''}\tilde{W}^{\scriptscriptstyle +-}({\bm q}_1-{\bm q}')
\psi_B^{(n)}({\bm K},{\bm q}')\nonumber\\
& &\times \psi_B^{(n)*}({\bm K},{\bm q}'')\tilde{W}^{\scriptscriptstyle +-}({\bm q}''-{\bm q}_2).
\end{eqnarray}
Since we are interested in a low momentum scattering, let us set ${\bm K}={\bm 0}$ and ${\bm q}_1={\bm q}_2={\bm 0}$. The low momentum scattering amplitude reads:
\begin{eqnarray}
\langle {\bm 0} {\bm 0}|\hat{T}(\epsilon)|{\bm 0} {\bm 0}\rangle
&=&\tilde{W}^{\scriptscriptstyle +-}({\bm 0})+
\sum_{n}\frac{g_{bx}^{(n)}({\bm 0},{\bm 0},{\bm 0})}{\epsilon-\Omega_{\bm 0}^{(n)}}.
\label{eq:scattering_amp}
\end{eqnarray}
In the first order Born approximation, the scattering amplitude is approximated with the lowest order term as $\langle {\bm 0} {\bm 0}|\hat{T}(\epsilon)|{\bm 0} {\bm 0}\rangle \simeq \tilde{W}^{\scriptscriptstyle +-}({\bm 0})$.
\begin{figure}
\includegraphics[width=0.5\textwidth]{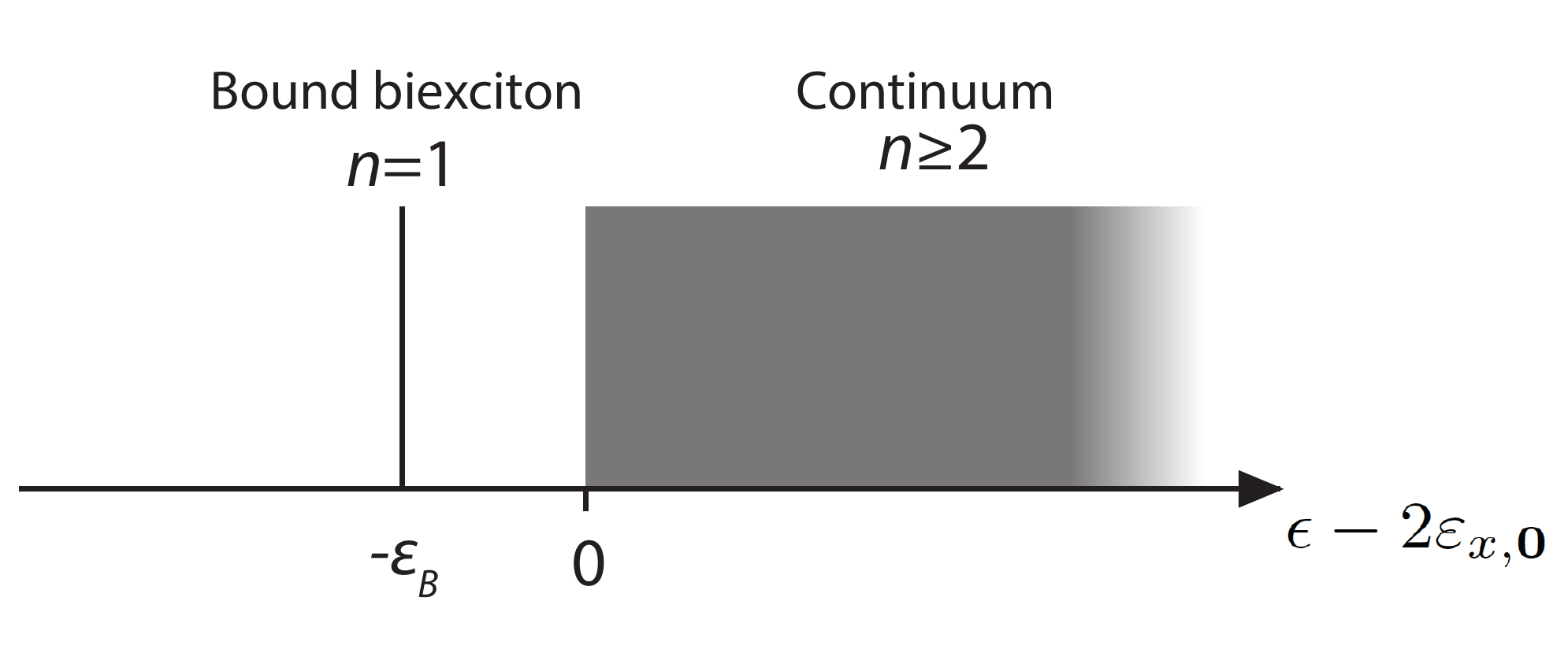}
\caption{Relative energy of the bound biexciton state and of the scattering continuum states.}
\label{fig:continuum}
\end{figure}

Now, let us consider the exciton-exciton interaction with anti-parallel spins. Since the exchange interaction does not exist at zero momentum, $\tilde{W}^{\scriptscriptstyle +-}({\bm 0})$ vanishes \cite{Ciuti1998,Inoue2000}. Thus, within the first order Born approximation, excitons with anti-parallel spins do not interact. However, based on the result of Ref. \cite{Takayama2002}, we will show that excitons with anti-parallel spins do interact due to the higher order contributions from the $T$-matrix expression. First, let us examine the structure of the two-exciton states. In general, the two-exciton states are composed of molecular bound states and unbound continuum states (also referred to as unbound biexcitons) \cite{Rodberg1967,Combescot1988,Oka2008a}. The energy of the bound states is lower than that of two free excitons. On the other hand, the energy of the unbound continuum states is higher than that of the two free excitons, and the number of unbound states is infinite. In the case of GaAs quantum well, it is known that the bare exciton-exciton interaction potential $\tilde{W}^{\scriptscriptstyle +-}({\bm q})$ includes one molecular bound state, which is nothing but the biexciton state considered in the main text. In addition to the biexciton, several calculations and experiments proved that the bare exciton-exciton interaction contains unbound continuum states, which is called ``scattering continuum" \cite{Axt1998,Kwong2001a}. Thus, we assume that the $n=1$ state is the bound biexciton state and its energy $\Omega_{\bm 0}^{(1)}$ is written as $\varepsilon_B$. Now the scattering amplitude Eq. \ref{eq:scattering_amp} reads as: 
\begin{eqnarray}
\langle {\bm 0} {\bm 0}|\hat{T}(\epsilon)|{\bm 0} {\bm 0}\rangle\simeq\frac{g_{bx}^{(1)}({\bm 0},{\bm 0},{\bm 0})}{\epsilon-\varepsilon_B}+
\sum_{n=2}^{\infty}\frac{g_{bx}^{(n)}({\bm 0},{\bm 0},{\bm 0})}{\epsilon-\Omega_{\bm 0}^{(n)}}.
\label{scattering-amplitude}
\end{eqnarray}
The structure of the eigenenergies of the $T$-matrix is schematically depicted in Fig. \ref{fig:continuum}. This is the main result of this appendix. In the effective exciton-exciton interaction Hamiltonian, the first part corresponds to the biexciton state and exciton-bexciton coupling. Meanwhile, the second part is the contribution to the scattering amplitude from the exciton continuum. As Eq. \ref{scattering-amplitude}  shows, the exciton continuum contribution depends on the energies of the incident excitons, however, the $\Omega_{\bm 0}^{(n)}$ are larger than the two free excitons energy and the energy dependence is not very sensitive, which is supported by the detailed calculation of the $T$-matrix in Ref. \cite{Takayama2002}. Therefore, it may be allowed to replace the exciton continuum contribution as a constant $g^{\scriptscriptstyle +-}$. In this appendix, we considered only the exciton-exciton interaction with anti-parallel spins. However, the same discussion can be applied to the exciton-exciton interaction with parallel spins $\tilde{W}^{\scriptscriptstyle ++}({\bm q})$ \cite{Kwong2001a}. Since the biexciton is not formed between excitons with parallel spins, the scattering amplitude is not very sensitive to the energy of two excitons and the scattering continuum works just as an offset to the the lowest order term $\tilde{W}^{\scriptscriptstyle ++}({\bm 0})$. The most important consequence of the scattering continuum
is appearance of the imaginary part of the T matrix. The imaginary part of the T matrix is equivalent to the energy dependent imaginary part of the interaction constant, which gives rise to the excitation induced dephasing (EID) term of the lower polaritons $\alpha'_1$ for the positive cavity detuning. 

\section*{Appendix C: Effective Hamiltonian of polariton-biexciton coupling}
 In the main text, the effective interaction constant $\alpha_2$ (Eq. \ref{eq:alpha2}) is indirectly obtained considering a pump-probe measurement. In this Appendix, we attempt to directly derive an effective Hamiltonian that represents the polaritonic Feshbach resonance but does not include biexciton operators. For this purpose, we make use of a canonical transformation called the Schrieffer-Wolff transformation \cite{Schrieffer1966,Taylor2002,Phillips2012,Cohen-Tannoudji1992}, which is originally introduced for Kondo physics but also has been applied to the derivation of the electron-phonon effective interaction Hamiltonian in superconductivity \cite{Frohlich1952,Taylor2002,Phillips2012}. In superconductivity, the electron-phonon interaction gives rise to an effective attractive interaction between electrons through a phonon retardation effect (Bardeen-Pines interaction\cite{Frohlich1952,Bardeen1955}). Actually, the polaritonic Feshbach resonance, where the polariton interaction is mediated by biexcitons, is analogous to the phonon mediated electron interaction. 

We consider a Hamiltonian representing lower polaritons, biexcitons, and the polariton-biexciton coupling in a momentum space $\hat{H}=\hat{H}_0+\hat{H}_I$:  
\begin{equation}
\hat{H}_0=\sum_{\sigma}\sum_{\bm k}\varepsilon_{L,\bm k}\hat{p}_{{\bm k},\sigma}^\dagger\hat{p}_{{\bm k},\sigma}+\sum_{\bm q}\varepsilon_{B,\bm q}\hat{m}_{\bm q}^\dagger\hat{m}_{\bm q}
\end{equation}
and
\begin{eqnarray}
\hat{H}_I=\sum_{\bm k \bm q}g^{bx}_{L}\Bigl[\hat{p}_{{\bm k},\uparrow}\hat{p}_{{- \bm k+ \bm q},\downarrow}\hat{m}_{\bm q}^\dagger+\hat{m}_{\bm q}\hat{p}_{{\bm k},\uparrow}^\dagger\hat{p}_{- \bm k+ \bm q,\downarrow}^\dagger\Bigr].
\end{eqnarray}
The Hamiltonian $\hat{H}_0$ and $\hat{H}_I$ correspond to the momentum space representations of the Hamiltonian Eq. \ref{eq:HlintLP} and Eq. \ref{eq:HintLP}. The annihilation operators of a lower polariton $\hat{p}_{\bm k,\sigma}$ and a biexciton $\hat{m}_{\bm k}$ are related to the field operators as  
\begin{equation}
\hat{\psi}_{L,\sigma}({\bf x})=\sum_{{\bm k}} \frac{1}{\sqrt{S}}e^{{\bm k}\cdot{\bf x}}\cdot\hat{p}_{\bm k,\sigma},
\end{equation}
and
\begin{equation}
\hat{\psi}_{B}({\bf x})=\sum_{{\bm k}} \frac{1}{\sqrt{S}}e^{{\bm k}\cdot{\bf x}}\cdot\hat{m}_{\bm k}.
\end{equation}
The energies $\varepsilon_{L,\bm k}$ and $\varepsilon_{B,\bm k}$ are given by
\begin{equation}
\varepsilon_{L(B),\bm k}=\varepsilon_{L(B)}+\frac{\hbar^2{\bm k}^2}{2m_{L(B)}}.
\end{equation}      
Now, we consider a canonical transformation of the type:  
\begin{eqnarray}
\hat{H}'&=&e^{-\hat{S}}\hat{H}e^{\hat{S}}\nonumber\\
&=&\hat{H}+[\hat{H},\hat{S}]+\frac{1}{2!}[[\hat{H},\hat{S}],\hat{S}]+...\ .
\label{eq:SW1}
\end{eqnarray}
When we chose the operator $\hat{S}$ such as to satisfy a relation: 
\begin{equation}
[\hat{H}_0,\hat{S}]=-\hat{H}_I,
\label{eq:SW_condition}
\end{equation}
With this expression, the Hamiltonian given by Eq. \ref{eq:SW1} is rewritten as
\begin{eqnarray}
\hat{H}'&=&e^{-\hat{S}}\hat{H}e^{\hat{S}}=\hat{H}_0+\frac{1}{2}[\hat{H}_I,\hat{S}]+...\ .
\end{eqnarray}
Here, we define an effective Hamiltonian $\hat{H}_{\rm eff}$ as  
\begin{eqnarray}
\hat{H}_{\rm eff}\equiv\frac{1}{2}[\hat{H}_I,\hat{S}].
\label{eq:Eeff}
\end{eqnarray}
In order to apply the canonical transformation to our Hamiltonian, we assume the operator $\hat{S}$ of the form\cite{Taylor2002,Phillips2012}: 
\begin{eqnarray}
\hat{S}&=&\sum_{\bm k \bm q}g^{bx}_{L}\Bigl[A_{{\bm k},{\bm q}}\hat{p}_{{\bm k},\uparrow}\hat{p}_{{- \bm k+ \bm q},\downarrow}\hat{m}_{\bm q}^\dagger+B_{{\bm k},{\bm q}}\hat{m}_{\bm q}\hat{p}_{{\bm k},\uparrow}^\dagger\hat{p}_{- \bm k+ \bm q,\downarrow}^\dagger\Bigr],\nonumber\\
\end{eqnarray}
\begin{figure}
\includegraphics[width=0.5\textwidth]{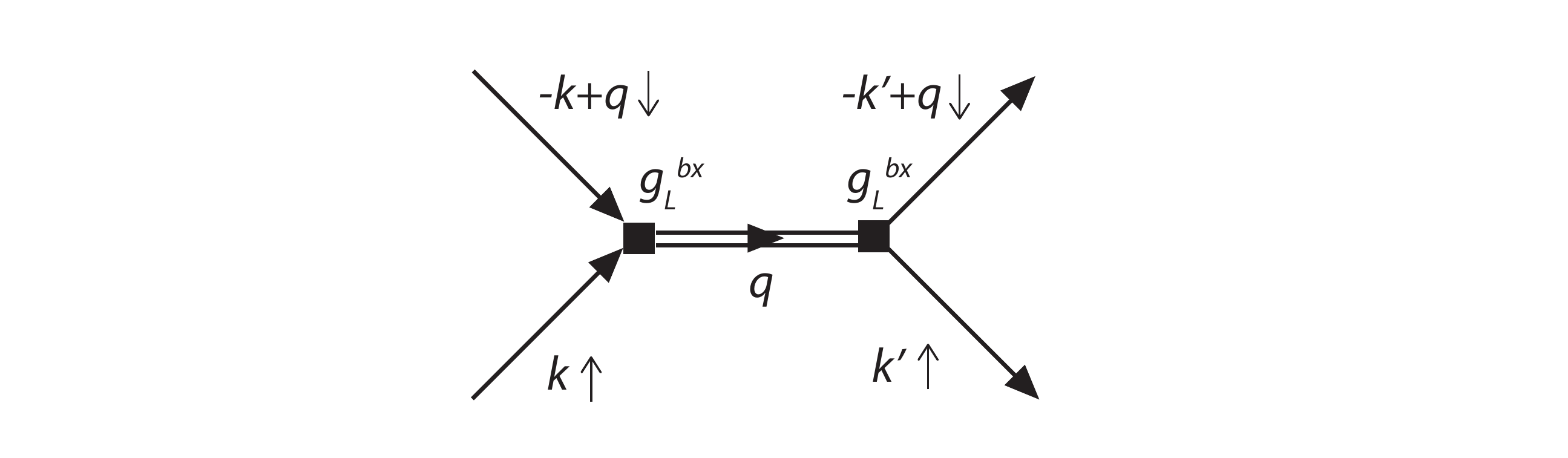}
\caption{Diagrammatic representation of the biexciton mediated effective polariton interaction, which is a second-order process. The arrow and the double line respectively represent a lower polariton and a biexciton}
\label{fig:SW_diagram}
\end{figure}\noindent where the coefficient $A_{{\bm k},{\bm q}}$ and $B_{{\bm k},{\bm q}}$ are determined such as to satisfy the condition Eq. \ref{eq:SW_condition}. We find the quantity $[\hat{H}_0,\hat{S}]$ as
\begin{eqnarray}
& &[\hat{H}_0,\hat{S}]=\nonumber\\
& &\sum_{\bm k \bm q}g^{bx}_{L}\Bigl[A_{{\bm k},{\bm q}}\left(-\varepsilon_{L,{\bm k}}-\varepsilon_{L,{-\bm k+\bm q}}+\varepsilon_{B,\bm q}\right)\hat{p}_{{\bm k},\uparrow}\hat{p}_{{- \bm k+ \bm q},\downarrow}\hat{m}_{\bm q}^\dagger\nonumber\\
& &+B_{{\bm k},{\bm q}}\left(\varepsilon_{L,{\bm k}}+\varepsilon_{L,{-\bm k+\bm q}}-\varepsilon_{B,\bm q}\right)\hat{m}_{\bm q}\hat{p}_{{\bm k},\uparrow}^\dagger\hat{p}_{- \bm k+ \bm q,\downarrow}^\dagger\Bigr].
\end{eqnarray}
In order to make this quantity equal to $-\hat{H}_0$, the coefficients $A_{{\bm k},{\bm q}}$ and $B_{{\bm k},{\bm q}}$ should be set as 
\begin{equation}
A_{{\bm k},{\bm q}}=\frac{1}{\varepsilon_{L,{\bm k}}+\varepsilon_{L,{-\bm k+\bm q}}-\varepsilon_{B,\bm q}}=\frac{1}{\Delta_{{\bm k},{\bm q}}}
\end{equation}
and
\begin{equation}
B_{{\bm k},{\bm q}}=\frac{1}{-\varepsilon_{L,{\bm k}}-\varepsilon_{L,{-\bm k+\bm q}}+\varepsilon_{B,\bm q}}=-\frac{1}{\Delta_{{\bm k},{\bm q}}},
\end{equation}
where $\Delta_{{\bm k},{\bm q}}$ is defined as
\begin{equation}
\Delta_{{\bm k},{\bm q}}=\varepsilon_{L,{\bm k}}+\varepsilon_{L,{-\bm k+\bm q}}-\varepsilon_{B,\bm q}.
\end{equation}
Now, we find that the operator for the canonical transformation $\hat{S}$ is given by
\begin{eqnarray}
\hat{S}&=&\sum_{\bm k \bm q}g^{bx}_{L}\Bigl[\frac{\hat{p}_{{\bm k},\uparrow}\hat{p}_{{- \bm k+ \bm q},\downarrow}\hat{m}_{\bm q}^\dagger}{\Delta_{{\bm k},{\bm q}}}-\frac{\hat{m}_{\bm q}\hat{p}_{{\bm k},\uparrow}^\dagger\hat{p}_{- \bm k+ \bm q,\downarrow}^\dagger}{\Delta_{{\bm k},{\bm q}}}\Bigr].\nonumber\\
\end{eqnarray}
Substituting this into Eq. \ref{eq:Eeff}, we obtain the effective Hamiltonian as
\begin{eqnarray}
\hat{H}_{\rm eff}&=&\frac{1}{2}[\hat{H}_I,\hat{S}]\nonumber\\
&=&\frac{1}{2}\sum_{\bm k \bm k' \bm q}g^{bx 2}_{L}\Bigl[\frac{1}{\Delta_{{\bm k'},{\bm q}}}+\frac{1}{\Delta_{{\bm k},{\bm q}}}\Bigr]\nonumber\\
& &\times\hat{p}_{{\bm k'},\uparrow}^\dagger\hat{p}_{- \bm k'+ \bm q,\downarrow}^\dagger\hat{p}_{{- \bm k+ \bm q},\downarrow}\hat{p}_{{\bm k},\uparrow}+O(\hat{p}^\dagger\hat{p}),
\end{eqnarray}
where the symbol $O(\hat{p}^\dagger\hat{p})$ represents terms that include only two lower polariton operators $\hat{p}^\dagger$ and $\hat{p}$. Since we are interested only in low momentum interactions (${\bm k}\sim{\bm k'}\sim{\bm q}\sim{\bm 0}$), using the approximation:
\begin{equation}
\Delta_{{\bm k},{\bm q}}\simeq\Delta_{{\bm k'},{\bm q}}\simeq 2\varepsilon_{L}-\varepsilon_{B},
\end{equation}
the effective Hamiltonian is rewritten as
\begin{eqnarray}
\hat{H}_{\rm eff}&\simeq&\sum_{\bm k \bm k' \bm q}\left(\frac{g^{bx 2}_{L}}{2\varepsilon_{L}-\varepsilon_{B}}\right)\hat{p}_{{\bm k'},\uparrow}^\dagger\hat{p}_{- \bm k'+ \bm q,\downarrow}^\dagger\hat{p}_{{- \bm k+ \bm q},\downarrow}\hat{p}_{{\bm k},\uparrow}\nonumber\\
&=&\left(\frac{g^{bx 2}_{L}}{2\varepsilon_{L}-\varepsilon_{B}}\right)\int d{\bf x}\ \hat{\bm \psi}_{L,\uparrow}^{\dagger}\hat{\bm \psi}_{L,\downarrow}^{\dagger}\hat{\bm \psi}_{L,\downarrow}\hat{\bm \psi}_{L,\uparrow}.
\end{eqnarray}
The prefactor of this effective Hamiltonian is same as the resonance part of the coefficient $\alpha_2$ (Eq. \ref{eq:alpha2}) with $\gamma_B=0$. This effective Hamiltonian is equivalent to the second-order perturbation and the process is diagrammatically presented in Fig. \ref{fig:SW_diagram}. The diagram in Fig. \ref{fig:SW_diagram} describes the biexciton mediated effective lower polariton interaction . The polariton energy dependence of the effective Hamiltonian is due to the retardation effect of biexcitons. The expression of the effective Hamiltonian is advantageous: since the Hamiltonian includes only lower polariton operators, the degree of freedom of biexciton is eliminated.

\end{document}